\documentclass[preprint,superscriptaddress,a4paper,showpacs,showkeys,11pt,nofootinbib]{revtex4}

\usepackage{amsmath}
\usepackage{amssymb}
\usepackage{graphicx}
\usepackage{hyperref}
\usepackage{xcolor}
\usepackage{booktabs}

\usepackage{geometry}
\usepackage{setspace}
\usepackage{booktabs}
\usepackage{float}
\usepackage{subfig}
\usepackage{hyperref}

\begin{document}

\title{Effects of QCD Equation of State on  Cosmological Parameters and  Stochastic Gravitational-Waves Spectrum}

\author{Andr\'e V. Giannini}
\email{AndreGiannini@ufgd.edu.br}
\affiliation{Faculty of Exact Sciences and Technology,
Federal University of Grande Dourados (UFGD), Postal code 364, 79804-970, Dourados, MS, Brazil}
\affiliation{Department of Physics, Santa Catarina State University, 89219-710 Joinville, SC, Brazil.}

\author{Victor P. Gon\c{c}alves}
\email{barros@ufpel.edu.br}
\affiliation{Institute of Physics and Mathematics, Federal University of Pelotas (UFPel), \\
  Postal Code 354,  96010-900, Pelotas, RS, Brazil}

\author{Samuel M. Sanches Jr.}
\email{samuelsanches@fei.edu.br}
\affiliation{Department of Physics, Centro Universitário FEI, \\ São Bernardo do Campo, SP, Brazil}
\affiliation{Centro Universitário FMABC, Santo André, SP, Brazil}

\begin{abstract}

We investigate the thermodynamic evolution of the early Universe across the QCD crossover and up to the electroweak (EW) scale, emphasizing the systematic impact of the QCD equation of state (EoS) and electroweak degrees of freedom on thermodynamic variables, cosmological parameters, and gravitational-wave observables. Using the HotQCD EoS as a baseline, we employ Gaussian Process (GP) regression to generate ensembles of 1000 thermodynamically consistent EoS samples, each extrapolated to high temperatures with and without EW contributions. For each ensemble, we solve the Friedmann equations to obtain the expansion history and compute gravitational-wave (GW) observables, including the instantaneous spectral tilt $n_{gw}(f)$ and the normalized GW energy density $\Omega_{gw}(f)/\Omega_{gw}^{\rm rad}$. We find that including the EW sector is the dominant effect across all observables, outweighing differences among QCD baseline models that share the same lattice calibration. In particular, EW contributions drive the EoS toward the radiation-dominated limit earlier, reduce excursions of the speed of sound toward the causality bound, suppress transient fluctuations in the Hubble and kinematic parameters near $t \sim 10^{-5}$~s, and qualitatively change the GW spectrum from a strongly frequency-dependent signal to a flat, suppressed plateau. These results demonstrate that accurate modeling of the electroweak sector is essential for an unbiased extraction of QCD-era physics from the stochastic gravitational-wave background.

\end{abstract}

\keywords{Early universe; Equation-of-state; Cosmological parameters; Gravitational waves.}

\date{\today}
\maketitle

\section{Introduction}
\label{sec:intro}
 
The thermal history of the Universe encodes a wealth of information about fundamental physics that is, in principle, accessible through cosmological observables \cite{Mazumdar:2018dfl}. At early times, the Universe was a hot, dense plasma of relativistic particles that expanded and cooled as it evolved, with the effective number of relativistic degrees of freedom changing across a sequence of phase transitions. One of these is the Quantum Chromodynamics (QCD) crossover, occurring at temperatures $T \sim 150$--$200$~MeV and times $t \sim 10^{-5}$~s, during which quarks and gluons confined into hadrons~\cite{Guenther:2020jwe}. The thermodynamic state of matter through this transition --- encoded in the equation of state (EoS) --- directly governs the expansion history of the Universe, which can be described in terms of observable cosmological scalar parameters (deceleration $q$, jerk $j$, \dots) defined by higher-order derivatives of the scale factor~\cite{Kolb:1990}. Moreover, the transition leaves imprints on a broad class of observables, from the light-element abundances produced during Big Bang nucleosynthesis to the stochastic gravitational-wave background (SGWB). In particular, SGWB of inflationary origin provide 
a  compelling observational window to the early Universe~\cite{Cai:2017cbj}. Primordial tensor perturbations generated during inflation re-enter the Hubble horizon throughout the radiation-dominated era; modes that re-enter during periods of non-conformal expansion acquire a frequency-dependent spectral tilt $n_{gw}(f) \neq 0$, carrying direct information about the EoS at the corresponding epoch~\cite{Schwarz:1997gv,Boyle:2007zx}. The QCD crossover, in particular, imprints a characteristic modulation in the GW spectrum at nanohertz-to-microhertz frequencies, a signal that may be within reach of pulsar-timing arrays and next-generation space-based detectors~\cite{Franciolini:2023wjm,Kawamura:2006up,Arzoumanian:2020vga,Amaro-Seoane:2017aoa}. Accurate predictions of this signal require precise knowledge of the EoS from the QCD scale up to, and beyond, the electroweak (EW) scale.
 
Significant progress has been made in lattice-QCD calculations of strongly interacting matter at zero baryon chemical potential. In particular, results from both the HotQCD~\cite{HotQCD} and Wuppertal--Budapest~\cite{Fodor:2014} collaborations are now well established and indicate that the transition between the quark--gluon plasma and a hadron gas is a crossover. These results also show that the pressure $P$ departs from its radiation-dominated limit, $P=\epsilon/3$, where $\epsilon$ is the energy density. This departure is quantified by the trace anomaly and induces a transient reduction of the expansion rate, which in turn affects the evolution of cosmological parameters and enhances the gravitational-wave signal. Overall, a realistic QCD EoS is essential for precise predictions of the early Universe's evolution.

Our goal in this paper is to extend these previous analyses by quantifying the impact of the QCD EoS on cosmological parameters and on the stochastic gravitational-wave spectrum. We combine a lattice-QCD-based equation of state with Gaussian-process (GP) regression to construct ensembles of 1000 EoS samples that systematically explore thermodynamic uncertainties across the QCD crossover. Building on the GP generative model introduced in~\cite{Gong:2025}, we extend each sample to high temperatures by adding the electroweak contribution as a relativistic gas with $g_\mathrm{ew}=14.45$ effective degrees of freedom, and we assess how the description of the QCD EoS affects a comprehensive set of cosmological observables. In particular, we estimate the temperature dependence of the energy density, pressure, and speed of sound, as well as the time evolution of the Hubble parameter and higher-order kinematic parameters ($q$, $j$) derived from the Friedmann equations. Finally, we quantify the impact of these EoS ensembles on the frequency dependence of gravitational-wave observables, including the equation-of-state parameter $\omega(T)$, the effective relativistic degrees of freedom $g_\mathrm{eff}(T)$, the instantaneous spectral tilt $n_{gw}(f)$, and the normalized GW energy density $\Omega_{gw}(f)/\Omega_{gw}^{\rm rad}$.

This paper is organized as follows.
Section~\ref{sec:framework} describes the theoretical framework, including the
baseline EoS, the GP generative model, the Friedmann equations used for
cosmological evolution, and the GW formalism.
Section~\ref{sec:results} presents the numerical results for cosmological parameters  and gravitational-wave observables. Finally, in
Section~\ref{sec:conclusions}, we summarize our main results and conclusions.

\section{Theoretical Framework}
\label{sec:framework}
 
\subsection{Baseline Equation of State}
\label{sec:eos}
\begin{figure}[t]
  \centering
  \includegraphics[width=0.7\columnwidth]{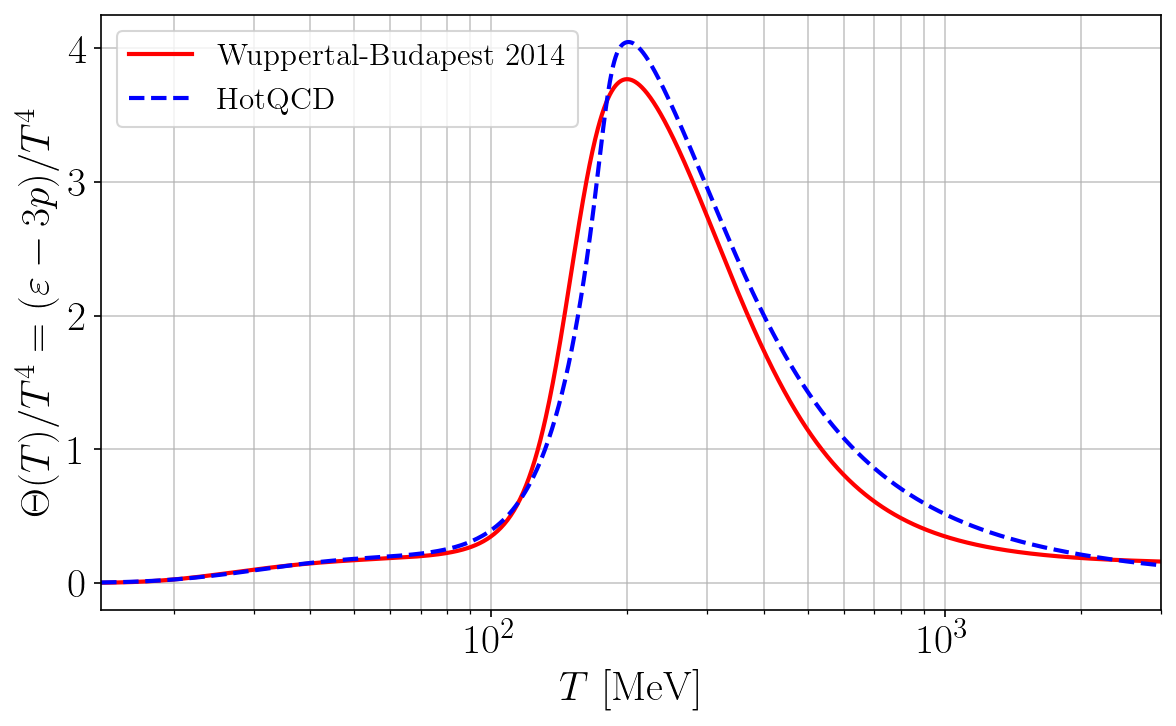}
  \caption{Trace anomaly $(\varepsilon - 3P)/T^4$ for the HotQCD and
           Wuppertal--Budapest EoS's as a function of temperature.}
  \label{fig:trace}
\end{figure}

The description of the transition from a quark--gluon plasma to a hadron gas has improved substantially over the past decades. In particular, the Wuppertal--Budapest~\cite{Fodor:2014} and HotQCD~\cite{HotQCD} collaborations have established that the QCD transition is a crossover, occurring at temperatures $T\sim150$--$200$~MeV. Both collaborations determine their EoS parameters using lattice-QCD data at zero baryon chemical potential. The remaining differences between the two EoS are subtle, and primarily reflect residual discretization and extrapolation uncertainties rather than distinct physics. In Fig.~\ref{fig:trace}, we compare their predictions for the trace anomaly,
\begin{equation}
  \frac{\Theta^{\mu}{}_{\mu}}{T^4}
  = \frac{\varepsilon - 3P}{T^4}\,,
  \label{eq:trace}
\end{equation}
where $\Theta^{\mu}{}_{\mu}$ is the trace of the energy--momentum tensor. A nonzero trace anomaly quantifies deviations from conformal behavior and, in particular, implies $P\neq\varepsilon/3$ through the crossover. Relative to the Wuppertal--Budapest prediction, the HotQCD result exhibits a peak at slightly higher temperature and with a larger maximum value. In what follows, we adopt the HotQCD EoS as our baseline.

\begin{figure}[t]
  \centering
  \includegraphics[width=\columnwidth]{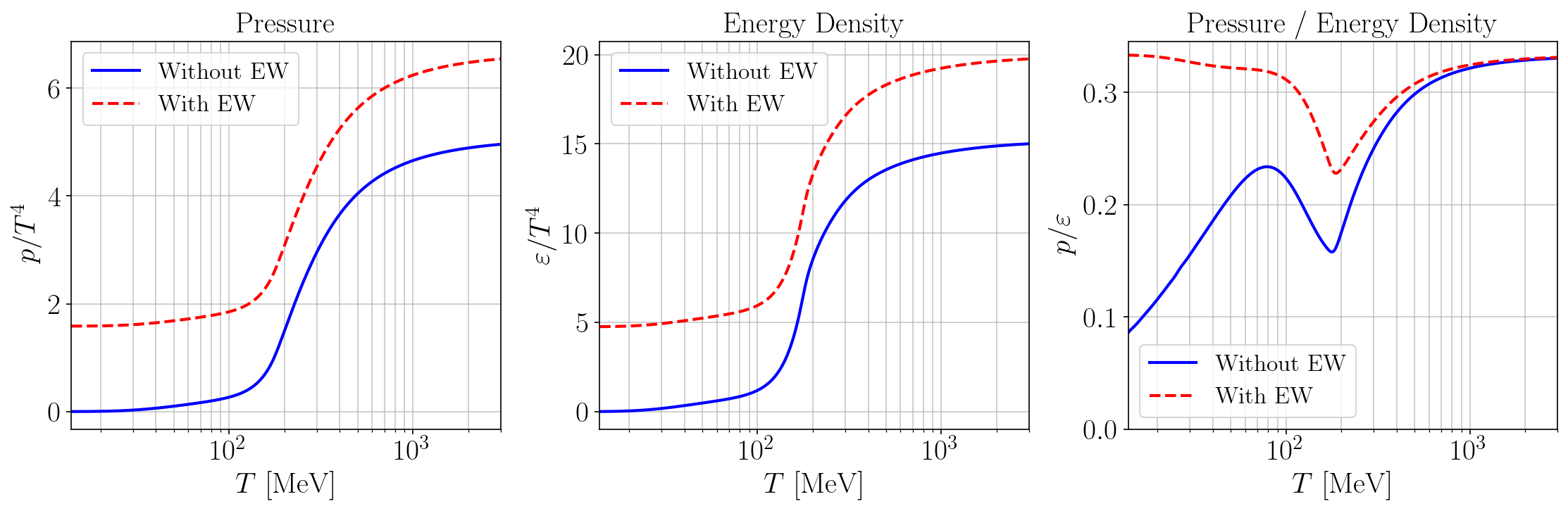}
  \caption{Pressure, energy density, and their ratio $P/\varepsilon$ for the
           HotQCD EoS with (dashed) and without (solid) the EW contribution.}
  \label{fig:eos_ew}
\end{figure}

In addition to the QCD sector, we include the electroweak contribution as a relativistic gas of effectively massless particles. Its energy density and pressure are~\cite{EWSector}
\begin{equation}
  \varepsilon_\mathrm{ew} = 3 P_\mathrm{ew}
  = g_\mathrm{ew}\,\frac{\pi^2}{30}\,T^4,
  \label{eq:ew}
\end{equation}
where $g_\mathrm{ew}=14.45$ is the effective number of electroweak degrees of freedom. The total energy density and pressure are then
\begin{equation}
  \varepsilon_T = \varepsilon_\mathrm{QCD} + \varepsilon_\mathrm{ew},
  \qquad
  P_T = P_\mathrm{QCD} + P_\mathrm{ew}.
  \label{eq:total}
\end{equation}
The impact of adding the electroweak contribution on the temperature dependence of thermodynamic observables is shown in Fig.~\ref{fig:eos_ew}. Both the energy density and pressure increase, while the ratio $P/\varepsilon$ is driven closer to $1/3$. This trend follows from the approximately conformal nature of the electroweak sector: since $P_\mathrm{ew}=\varepsilon_\mathrm{ew}/3$ exactly in this approximation, any admixture of electroweak radiation pulls the total EoS toward the radiation-dominated limit. It follows trivially that the trace anomaly vanishes exactly for the EW sector, so results shown in Fig. \ref{fig:trace} do not change.

\subsection{Gaussian Process Generative Model}
\label{sec:gp}
 
To investigate cosmological implications without relying on a single
functional parametrization, we employ the Gaussian Process (GP) generative
model for the dimensionless scaled pressure $\ln(P/T^4)$ introduced
in~\cite{Gong:2025} (See also Ref.~\cite{Gong:2025data}). The GP is trained using the baseline QCD EoS in two
temperature windows, $T \leq 0.13$~GeV and $T \geq 0.7$~GeV, matching the
range of Ref.~\cite{Gong:2025}. After sampling, only EoS's that satisfy the
following thermodynamic and causality conditions for all temperatures are
retained:
\begin{equation}
  \frac{\partial P}{\partial T} > 0,
  \qquad
  \frac{\partial^2 P}{\partial T^2} > 0,
  \qquad
  0 \leq c_s^2 \left(\equiv \frac{\partial P}{\partial \varepsilon}\right) < c_{s,\mathrm{max}}^2.
  \label{eq:conditions}
\end{equation}
In this study we focus on the subset of samples satisfying
$c_{s,\mathrm{max}}^2 \leq 1.0$, which allows for significant deviations from
the ideal-gas limit $c_s^2 = 1/3$.
The GP is trained exclusively on QCD information. When applicable, the EW contribution is
added after sampling a QCD EoS from the trained GP but \emph{before} checking the thermodynamic and causality conditions~\eqref{eq:conditions}. For each baseline EoS, we
generate an ensemble of 1000 accepted samples. 

It is known that imposing constraints while sampling a GP generally modifies the underlying probability distribution of the resulting ensemble~\cite{Guimaraes:2026tae}. This, however, does not impact the investigation carried out here since our goal is to explore the space of possible EoS without assuming a particular functional parameterization, not to constrain it via a Bayesian analysis. Moreover, the ensemble is intentionally broader than the family of EoS compatible with existing Lattice QCD data. The main goal is to consider the present GP as a tool to test sensitivity and robustness of cosmological quantities of interest, in this way we can assess whether known signatures survive substantial changes in the input EoS. A weak dependence would indicate that a given observable is insensitive to the detailed shape of the crossover region, while a strong dependence could be traced back to the thermodynamic condition it would be affect by.
This question is relevant for both, interpreting the future gravitational-wave measurements and for examining how much information about the QCD EoS could, in principle, be extracted from them.

\begin{figure}[t]
	\centering
	\subfloat[HotQCD without EW contribution]{\includegraphics[width=0.5\textwidth]{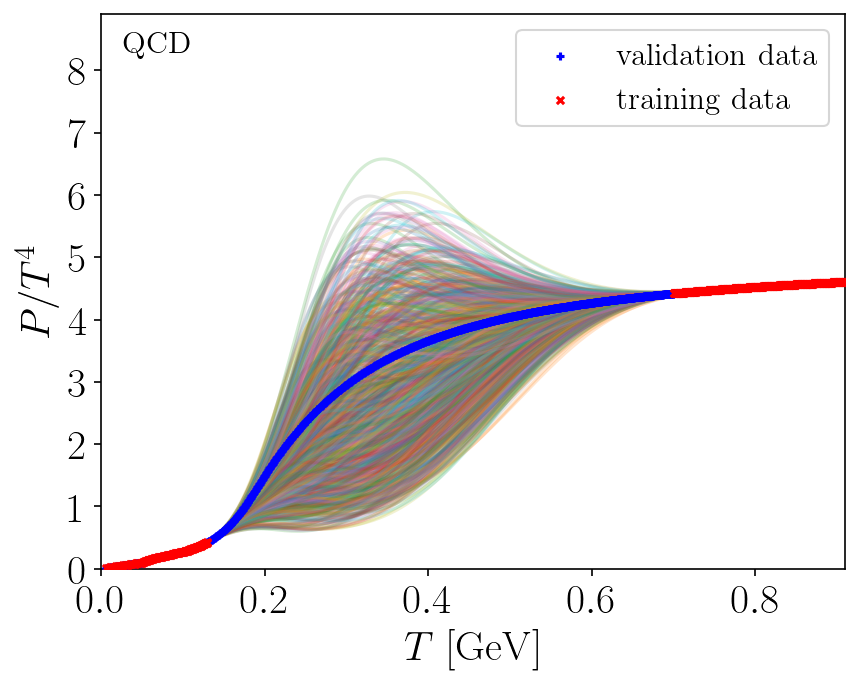}\label{fig:gp_pressureA}}
	\subfloat[HotQCD with EW contribution]{\includegraphics[width=0.5\textwidth]{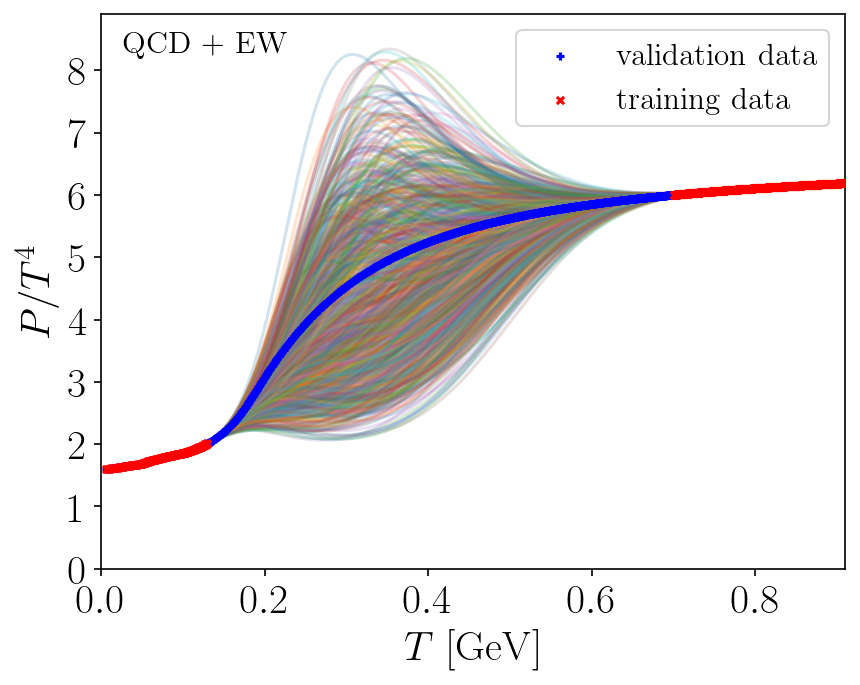}\label{fig:gp_pressureB}}
	\caption{$P/T^4$ from 1000 GP-generated EoS samples. Training data (crosses) and validation data (plus signs) are
           shown for reference.}
    \label{fig:gp_pressure}
\end{figure}

\begin{figure}[t]
	\centering
	\subfloat[without EW contribution]{\includegraphics[width=0.5\textwidth]{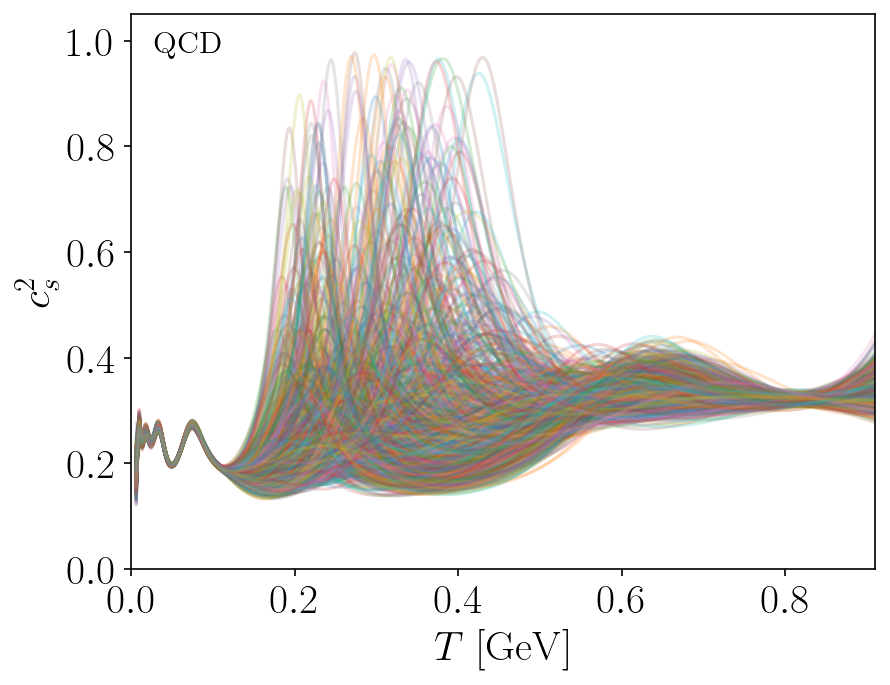}\label{fig:cs2A}}
    \subfloat[with EW contribution]{\includegraphics[width=0.5\textwidth]{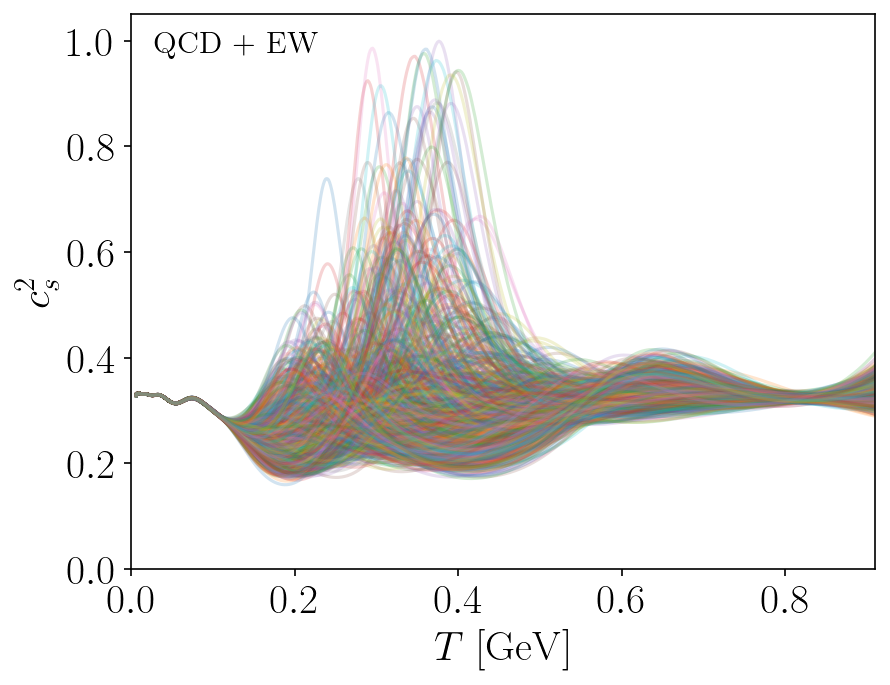}\label{fig:cs2B}}
	\caption{Speed of sound $c_s^2$ for each EoS generated via the Gaussian
           Process.}\label{fig:cs2}
\end{figure}

In Fig.~\ref{fig:gp_pressure} we present our results for the dimensionless scaled pressure $P/T^4$ considering the 1000 GP-generated samples, with and without the EW contribution. In the
QCD-only case (Fig.~\ref{fig:gp_pressureA}), the ensemble displays a visible
spread around the QCD crossover region, reflecting the GP uncertainty. Upon
adding the EW contribution (Fig.~\ref{fig:gp_pressureB}), the entire
distribution is shifted upward, and the relative impact of QCD-induced
fluctuations is reduced. 
This effect persists across the full ensemble of 1000 GP-generated EoS
samples.  This result is consistent with recent efforts to construct unified
Standard Model EoS across wide temperature
ranges~\cite{Laine:2006cp,Borsanyi:2016ksw}, which emphasize the importance
of including all relativistic degrees of freedom for accurate cosmological
predictions.

The corresponding speed of sound is shown in Figure~\ref{fig:cs2}. Without the
EW contribution, a significant fraction of GP samples approach the causality
bound $c_s^2 \to 1$ over a broad temperature interval
$T \sim 0.2$--$0.4$~GeV encompassing the QCD crossover. After including the
EW sector, this behavior is substantially reduced: excursions toward the
causality bound become less frequent and are confined to a narrower
temperature range. Physically, the conformal EW component dilutes the
non-conformal QCD features responsible for large gradients in
$\partial P/\partial\varepsilon$.
 This indicates that the EW sector acts as a thermodynamic regulator,
effectively damping large fluctuations in $\partial P/\partial\varepsilon$ and
reducing excursions toward the causality bound across the GP ensemble.
Physically, this can be understood as a consequence of adding a nearly
conformal relativistic component ($P \approx \varepsilon/3$), which dilutes
non-conformal QCD features. This behavior aligns with studies showing that
deviations from conformality in QCD can lead to nontrivial peaks in the speed
of sound under extreme conditions~\cite{HotQCD,Fodor:2014}. Our results
demonstrate that these effects are naturally mitigated when the EW sector is
added, even without modifying the QCD sector itself.

\subsection{Cosmological Parameters}
\label{sec:friedmann}

The expansion of a flat, radiation-dominated Universe is governed by the Friedmann equations, obtained by solving the Einstein field equations for a Friedmann--Lema\^{\i}tre--Robertson--Walker (FLRW) metric filled with a perfect fluid of total energy density $\varepsilon_T$ and pressure $P_T$ \cite{Weinberg:1972,Kolb:1990}. In standard form,
\begin{equation}
  H^2 = \left(\frac{\dot{a}}{a}\right)^2
  = \frac{8\pi G}{3}\,\varepsilon_T,
  \qquad
  \frac{\ddot{a}}{a}
  = -\frac{4\pi G}{3}\left(\varepsilon_T + 3P_T\right),
  \label{eq:friedmann}
\end{equation}
where $a(t)$ is the (dimensionless) scale factor, $\dot{a}$ and $\ddot{a}$ are its first and second derivatives with respect to cosmic time, and $H$ is the Hubble parameter, which quantifies the expansion rate. Here, $G$ denotes Newton's gravitational constant. To close the system one must specify an equation of state  for the relevant degrees of freedom.

In recent years, several authors have pursued an alternative approach to cosmic evolution based on kinematic parameters defined through higher-order derivatives of the scale factor. These are commonly referred to as cosmographic (or kinematic) models ~\cite{Visser:2004bf}. Their main motivation is that expansion parameters can be inferred more directly from data, with fewer assumptions than required in a dynamical modelling framework (See, e.g., Ref.~\cite{Jesus:2022xwb}). In our analysis, we will focus on the deceleration $q$ and jerk $j$ kinematical parameters, which are defined by
\begin{align}
  q &= -\frac{\ddot{a}}{aH^2}
       \quad \text{(deceleration)},
  \label{eq:q}\\
  j &= \frac{\dddot{a}}{aH^3}
       \quad \text{(jerk)}.
  \label{eq:j}
\end{align}
For a flat Universe with an equation of state $P_T = c_s^2\,\varepsilon_T$ and constant speed of sound $c_s$, the Friedmann equations imply \cite{Castorina:2018gsu}
\begin{equation}
  q = \frac{1 + 3c_s^2}{2},
  \qquad
  j = \left(\frac{1 + 3c_s^2}{2}\right)\left(2 + 3c_s^2\right)\,.
\end{equation}
Therefore, in a radiation-only Universe one finds $q = 1$ and $j = 3$. Departures from these values quantify the mixture of radiation and matter and/or the presence of interactions between the two components. As shown in the previous subsection, the speed of sound is not constant during the QCD transition; consequently, the cosmological parameters are expected to be sensitive to the QCD equation of state.

\subsection{Stochastic Gravitational-Waves}
\label{sec:gw_formalism}
Cosmological phase transitions are expected to modify the spectrum of gravitational waves generated during inflation by fluctuations of the metric tensor~\cite{Schwarz:1997gv,Boyle:2007zx}. Detecting this low-frequency stochastic gravitational-wave (SGW) background would improve our understanding of the Universe at very early times and help constrain the phase transitions it underwent during its evolution. Upcoming experiments like Pulsar Timing Array (PTA) \cite{Shaifullah:2025pwz} and Square Kilometer Array (SKA) \cite{Dewdney:2009tmd} are expected to probe the frequency range of SGW backgrounds. 

To estimate the present-day spectrum of the stochastic gravitational-wave (GW) background, one must evolve the signal from its production epoch to today. Here we focus on the impact of the QCD transition on the GW spectrum. Previous studies have shown that this transition produces a step-like feature, driven by the decrease in the number of relativistic degrees of freedom. During the transition, the rate at which modes enter the horizon changes, imprinting a step at frequencies of order the Hubble rate at the transition. In what follows, we present a brief review of the formalism needed to estimate the GW spectrum. For a more detailed discussion, we refer the interested reader to the Refs.~\cite{
Schettler:2010dp,Anand:2017kar,Castorina:2018gsu,Chen:2017cyc,Dai:2019ksi,Moussa:2021gxb,Franciolini:2023wjm}.

Assuming that the Universe has expanded adiabatically since the phase transition, the Friedmann equation implies that the time variation of the temperature is given by
\begin{eqnarray}
    \frac{dT}{dt} = - \frac{HT}{1 + \frac{T}{3g_\mathrm{eff}}\frac{dg_\mathrm{eff}}{dT}} \,\,,
    \label{eq:Temp}
\end{eqnarray}
where $g_\mathrm{eff}$ is the effective number of relativistic degrees of freedom that contributes to the entropy density, which can be defined estimated assuming that
\begin{equation}
  \varepsilon(T) = \frac{\pi^2}{30}\,g_\mathrm{eff}(T)\,T^4\,.
\end{equation}
The Eq. (\ref{eq:Temp}) allow us to estimate the ratio between the scale factor at the epoch of the transition $a_*$ and its value at the current epoch $a_0$, which is given by
\begin{eqnarray}
    \frac{a_*}{a_0} = \exp\left[\int_{T_*}^{T_0} \frac{1}{T}\left(1 + \frac{T}{3g_\mathrm{eff}}\frac{dg_\mathrm{eff}}{dT}\right) dT\right] \,\,.
\end{eqnarray}
As a consequence, the density parameter of the   gravitational waves at current epoch, $\Omega_{gw}$, can be expressed as follows
\begin{eqnarray}
    \Omega_{gw} = \Omega_{gw *}\left(\frac{H_*}{H_0}\right)^2 \exp\left[\int_{T_*}^{T_0} \frac{4}{T}\left(1 + \frac{T}{3g_\mathrm{eff}}\frac{dg_\mathrm{eff}}{dT}\right) dT\right]
\end{eqnarray}
The ratio between the Hubble parameter at the epoch of the transition to that of its value today can be estimated using the continuity equation, and is given by
\begin{eqnarray}
   \left(\frac{H_*}{H_0}\right)^2 = \Omega_{r0}\left(\frac{a_0}{a_r}\right)^4 \exp\left[\int_{T_r}^{T_*} \frac{3(1+\omega)}{T}\left(1 + \frac{T}{3g_\mathrm{eff}}\frac{dg_\mathrm{eff}}{dT}\right) dT\right] \,\,, 
\end{eqnarray}
where $\Omega_{r0}$ is the current value of the fractional energy density of radiation and $T_r$ is some early time in the radiation dominated epoch, characterized by the scale factor $a_r$, before the epoch of the QCD phase transition.
Moreover, $\omega(T) = P(T)/\varepsilon(T)$ is the equation-of-state parameter, which can be obtained directly from the EoS ensembles. As a consequence, the trace anomaly in the QCD EoS alter the fractional energy density of the GW.

The logarithmic slope of the GW energy spectrum encodes the equation of state
at horizon re-entry. Following Ref.~\cite{Boyle:2007zx}, deviations from conformality
imprint a characteristic tilt, given by
\begin{equation}
  n_{gw} \equiv \frac{d\ln\Omega_{gw}}{d\ln f} = \frac{2(3\omega - 1)}{1 + 3\omega}\,,
    \label{eq:ngw}
\end{equation}
where $f$ is the GW frequency. For a radiation dominated epoch, where  $\omega = 1/3$, $n_{gw} =0$. On the other hand, as during the QCD transition $\omega \neq 1/3$, such a quantity becomes sensitive to the EoS. Finally, Eq. (\ref{eq:ngw}) allow us to derive the frequency dependence of the fractional energy density by a simple integration.

\section{Results}
\label{sec:results}
In what follows, we present our results for thermodynamic variables, cosmological parameters, and stochastic GW observables, focusing on the impact of the QCD EoS on the temperature, time, and frequency dependencies of these quantities. To quantify the role of electroweak degrees of freedom, we also show results obtained by omitting them. The calculations are performed using the GP-generated ensemble of 1000 EoSs.

\begin{figure}[t]
	\centering
	\subfloat[Temperature, QCD]{\includegraphics[width=0.5\textwidth]{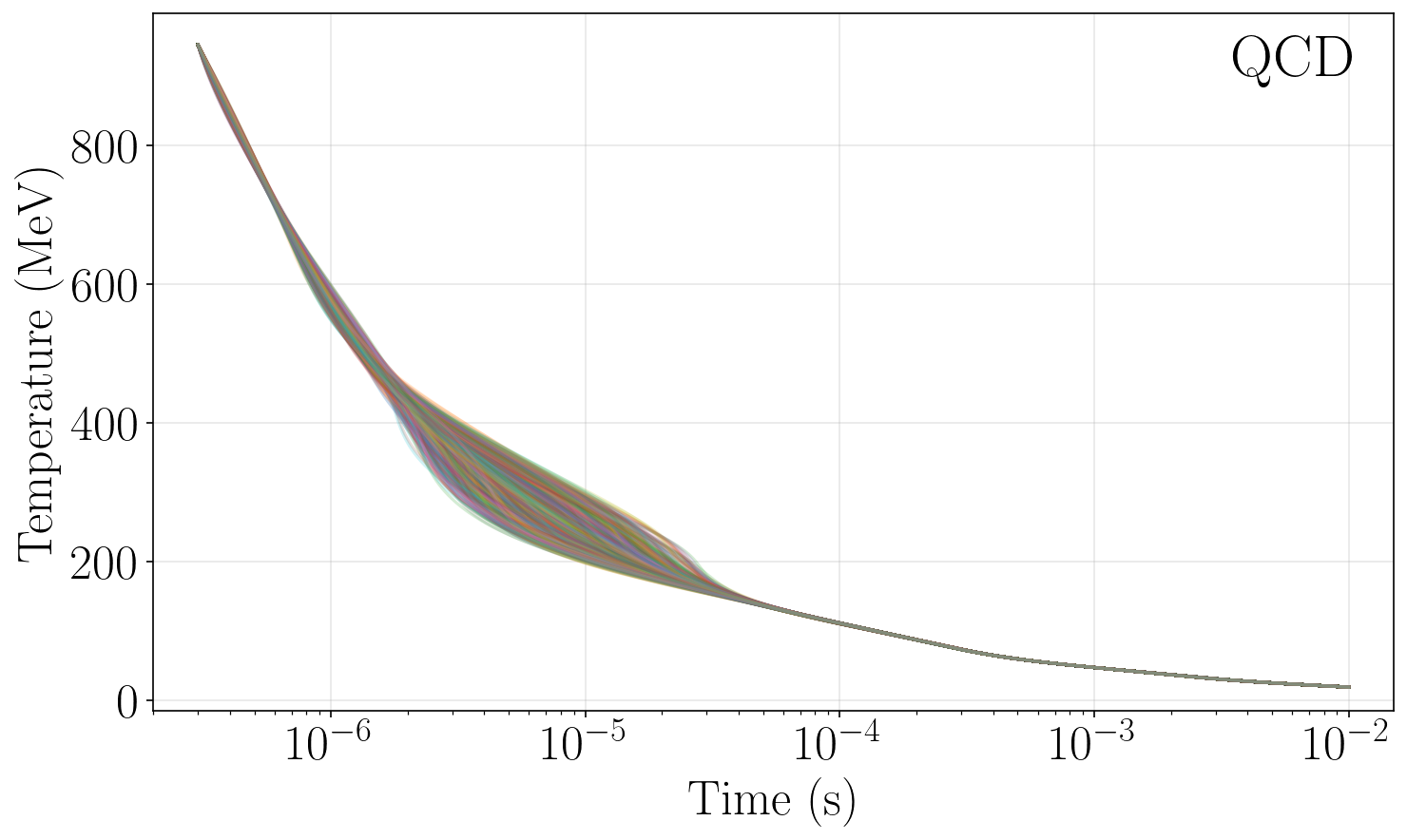}}
	\subfloat[Temperature, QCD + EW]{\includegraphics[width=0.5\textwidth]{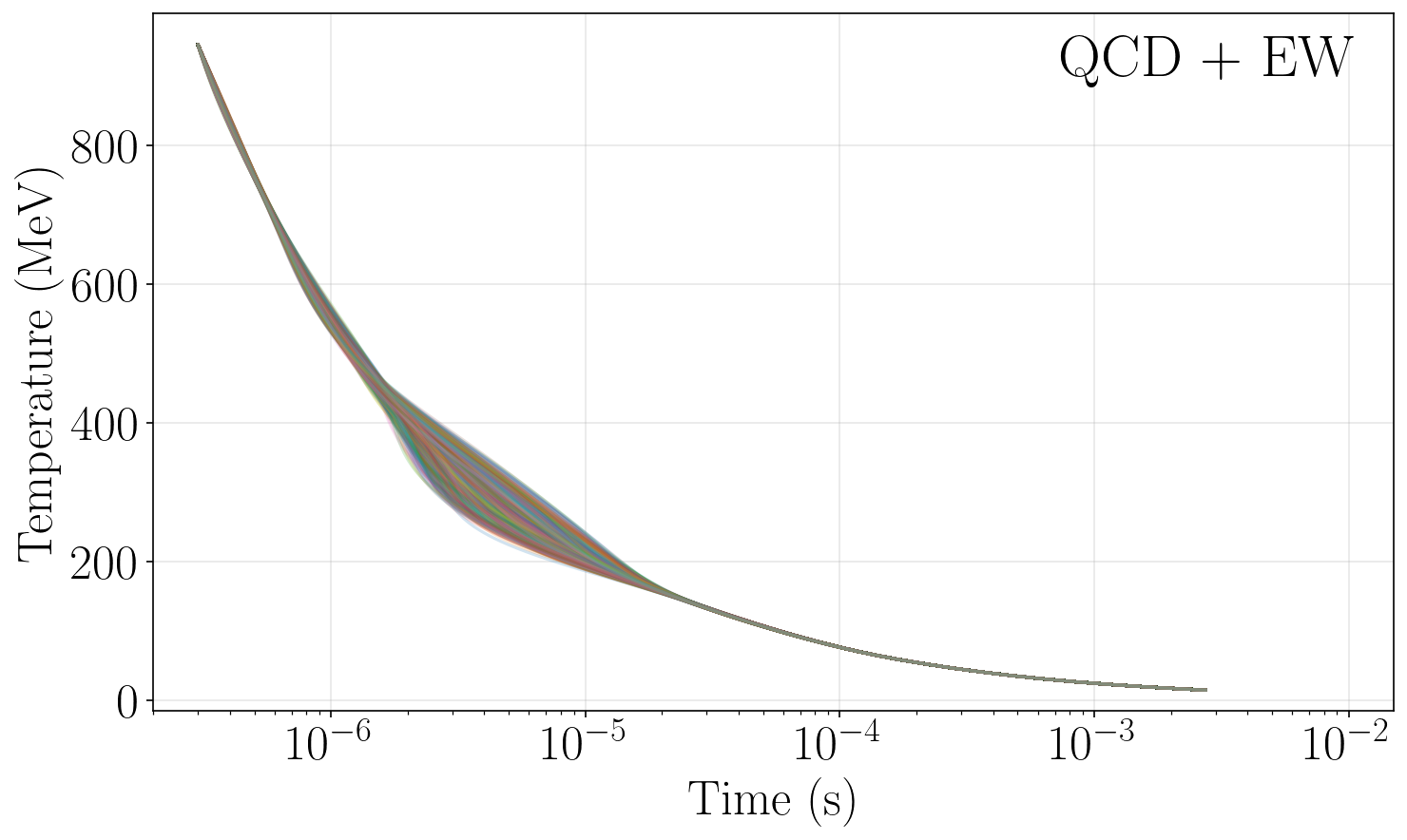}}\\
    \subfloat[$P/\varepsilon$, QCD]{\includegraphics[width=0.5\textwidth]{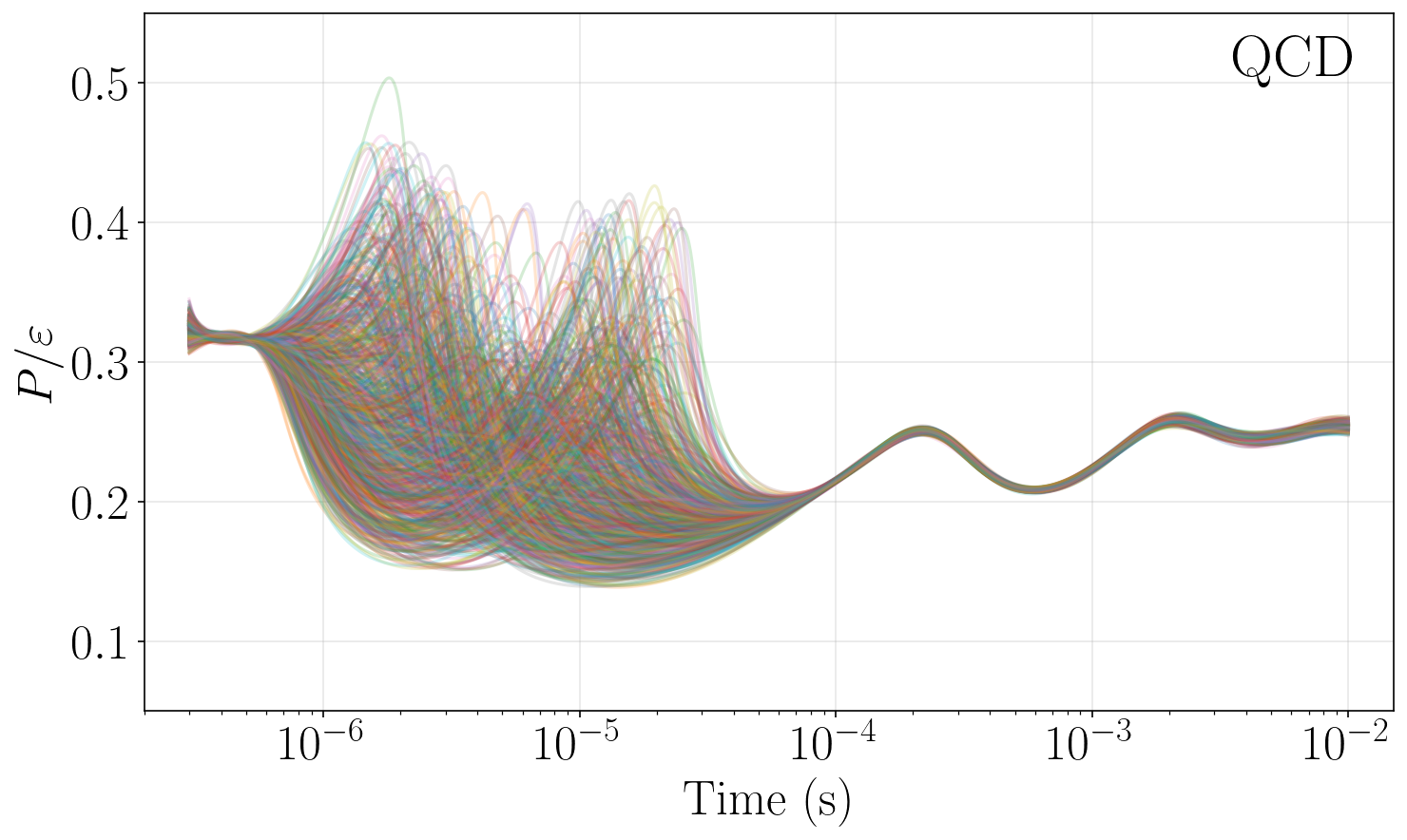}}
    \subfloat[$P/\varepsilon$, QCD + EW]{\includegraphics[width=0.5\textwidth]{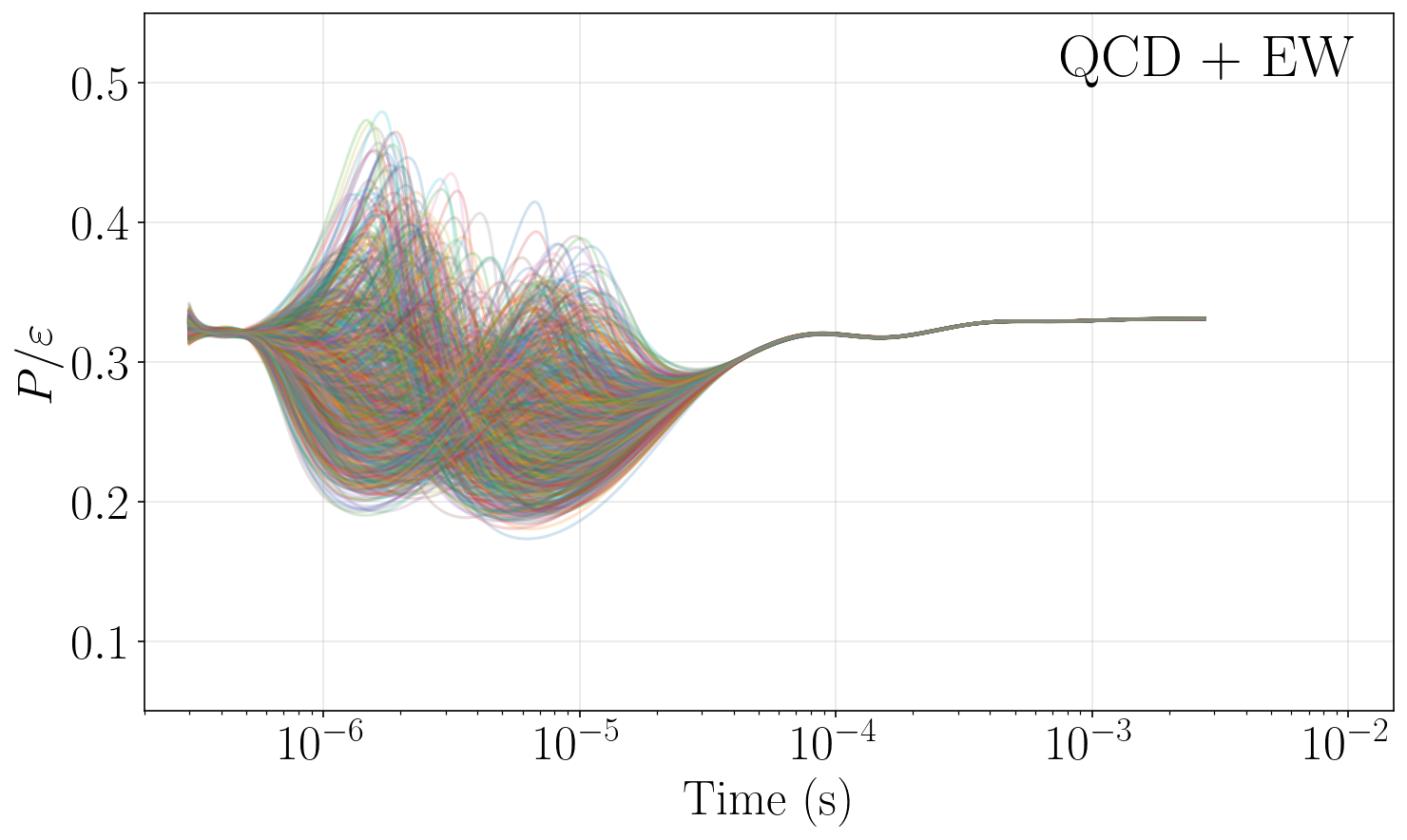}}
	\caption{Time evolution of thermodynamic quantities for 1000 GP-generated
           EoS.}
    \label{fig:thermo_time}
\end{figure}

\begin{figure}[t]
	\centering
    \subfloat[Hubble parameter, QCD]{\includegraphics[width=0.5\textwidth]{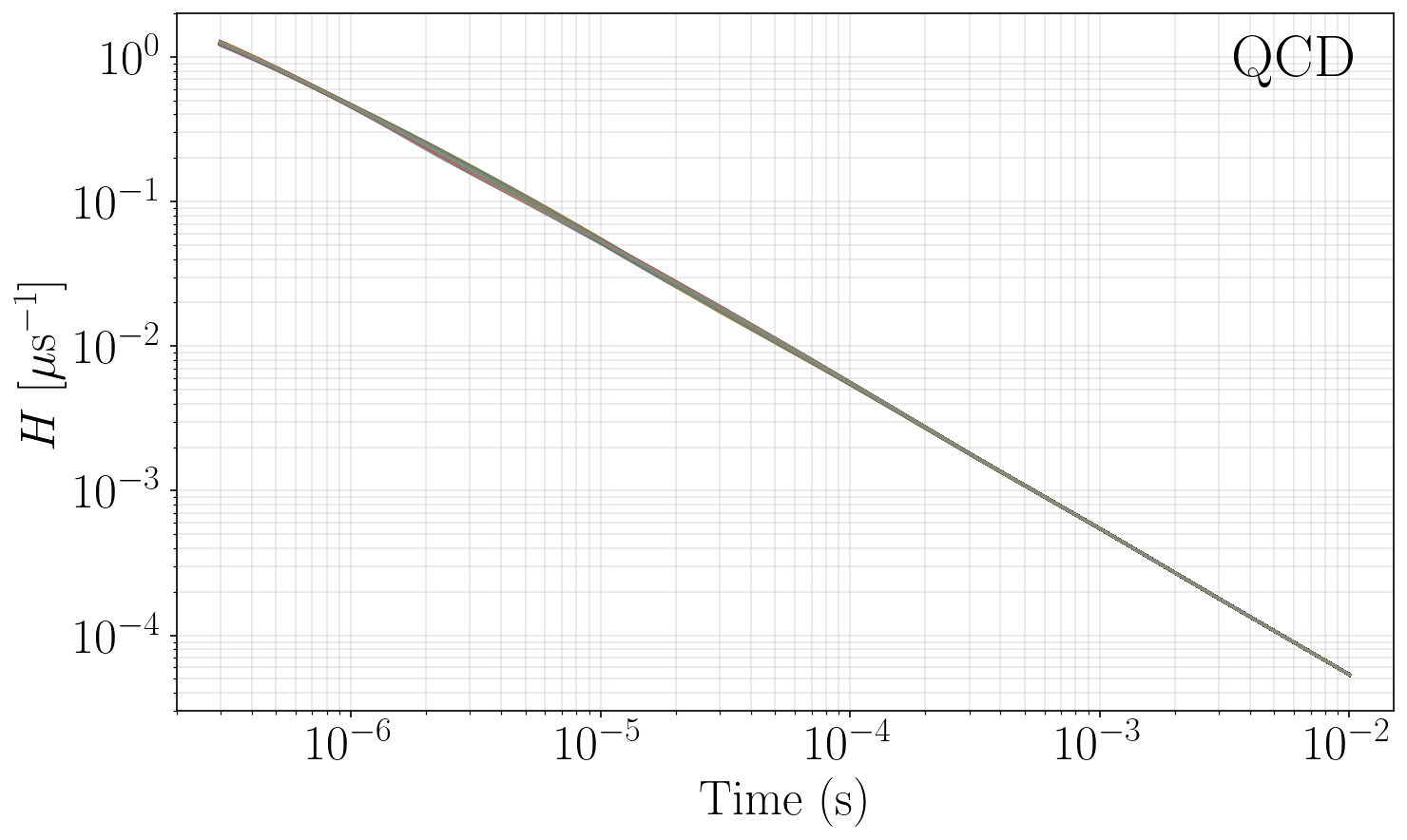}\label{fig:cosmo_paramsC}}
	\subfloat[Hubble parameter, QCD + EW]{\includegraphics[width=0.5\textwidth]{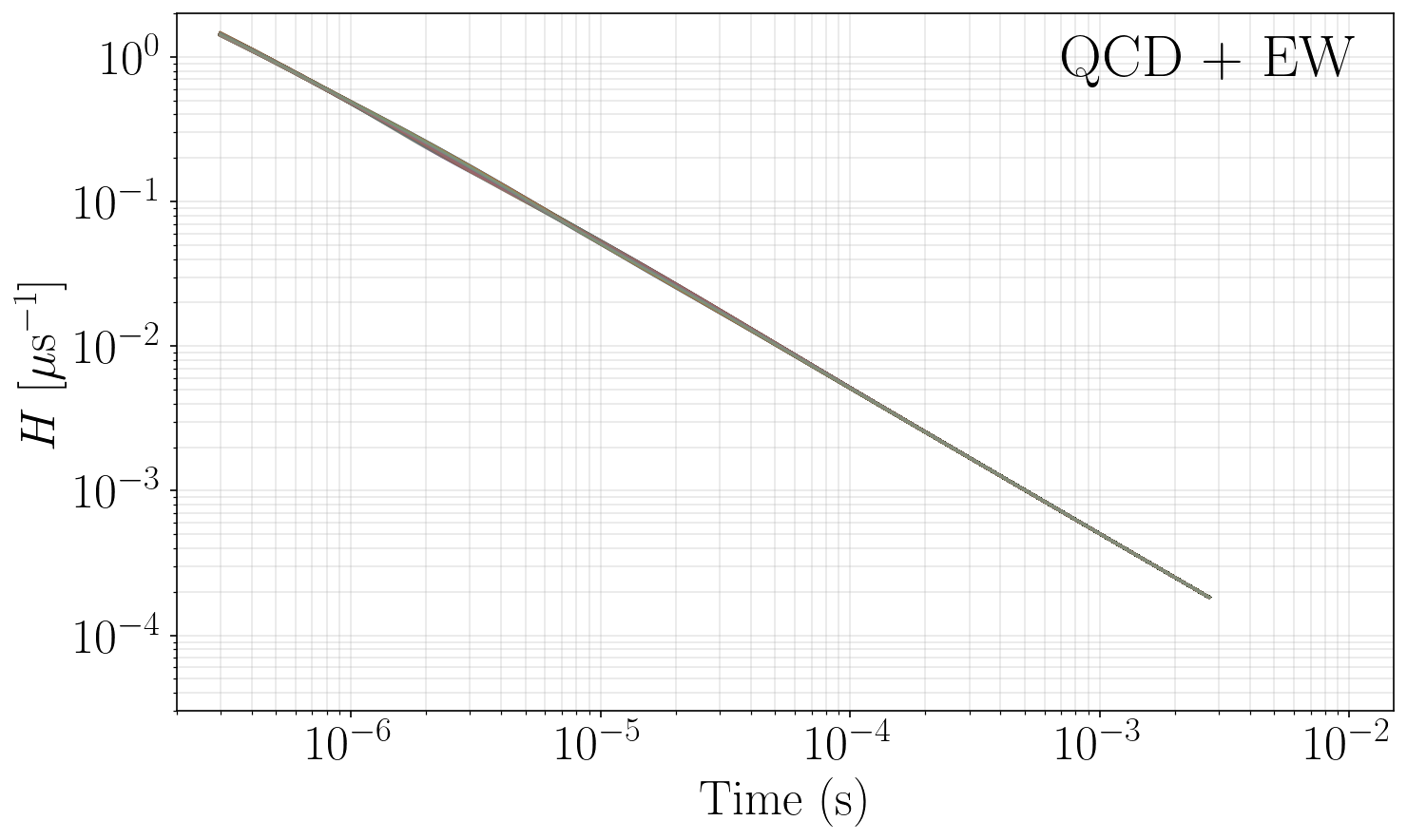}\label{fig:cosmo_paramsD}}\\
    \subfloat[Deceleration parameter,  QCD]{\includegraphics[width=0.5\textwidth]{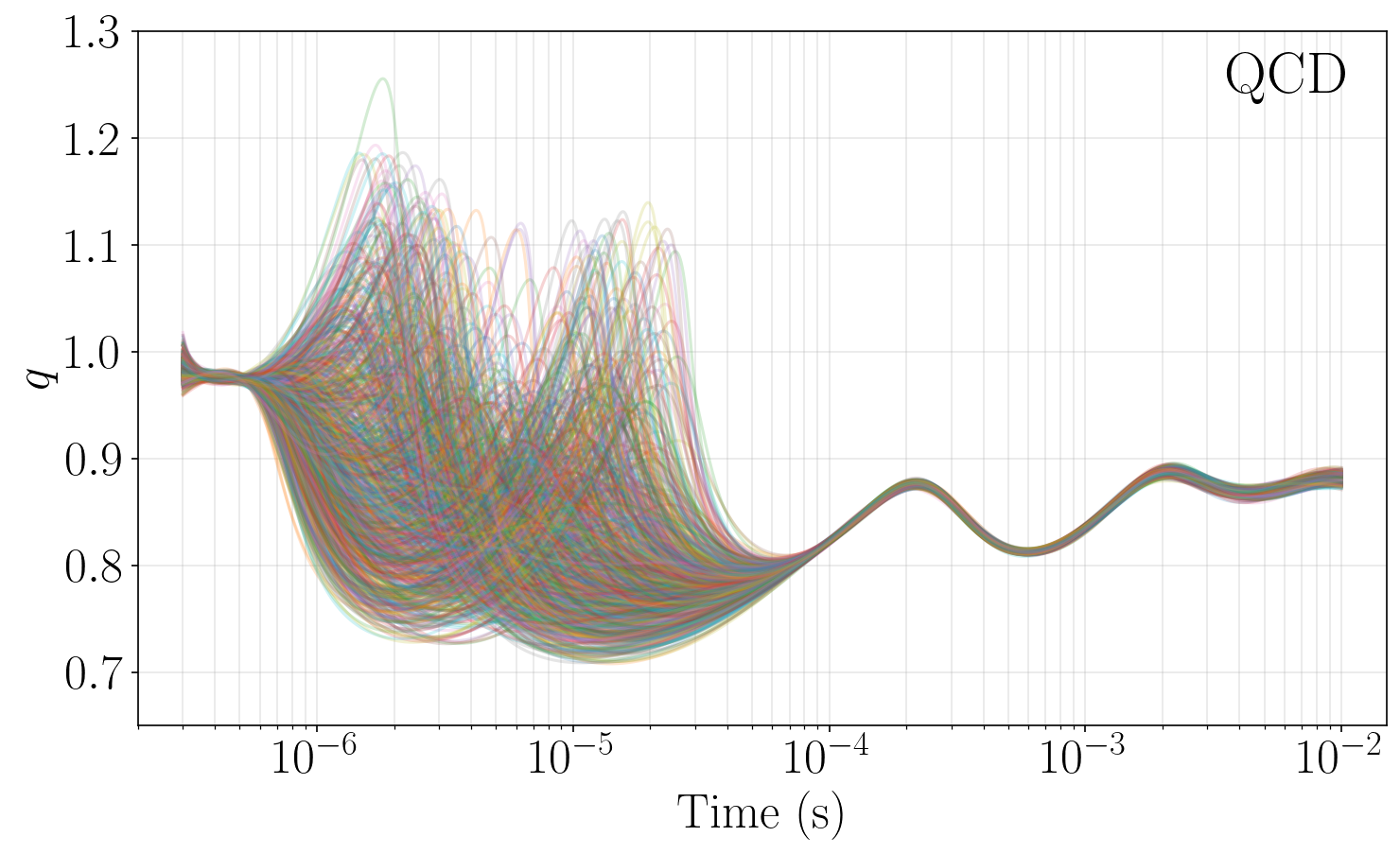}\label{fig:cosmo_paramsE}}
	\subfloat[Deceleration parameter, QCD + EW]{\includegraphics[width=0.5\textwidth]{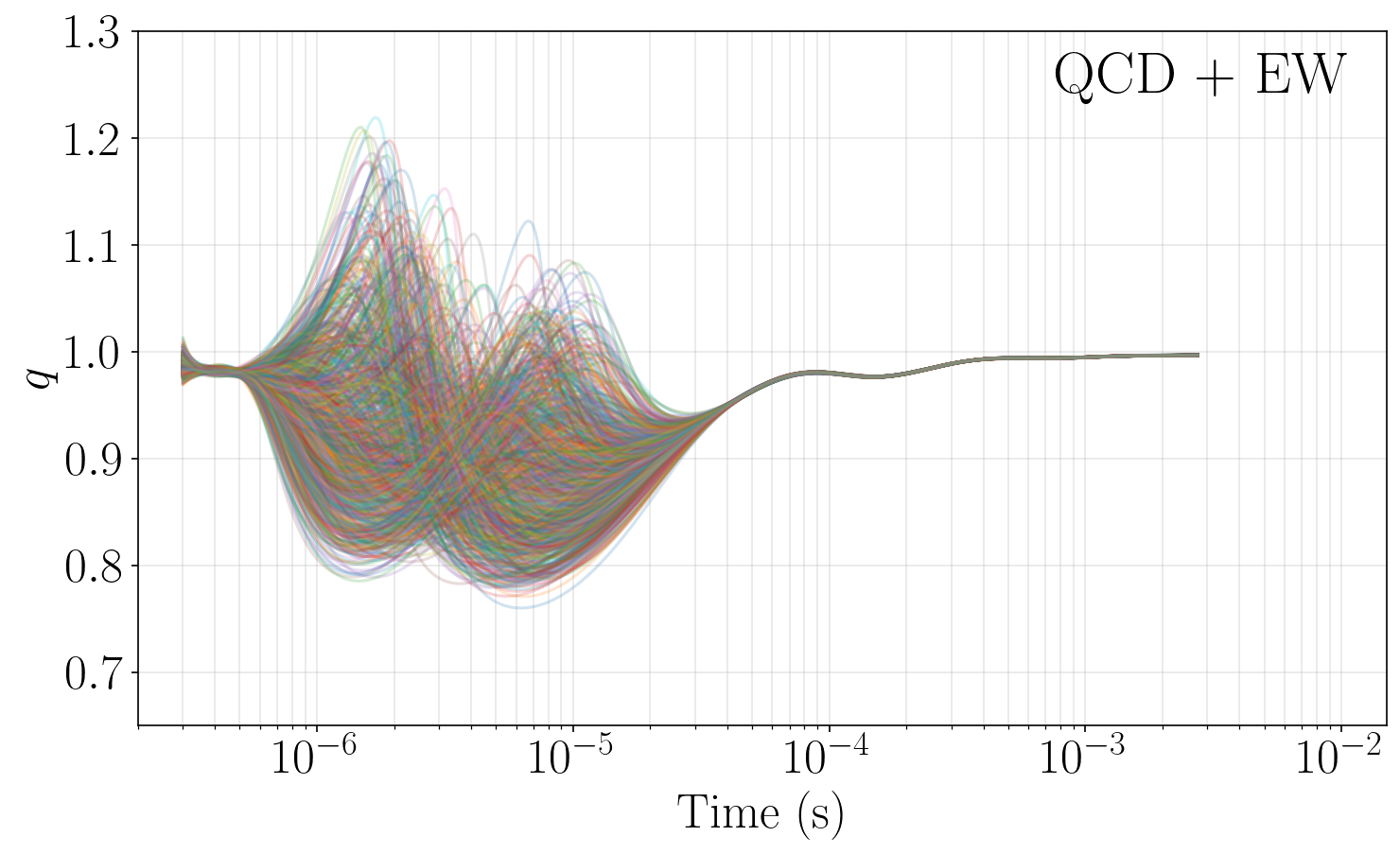}\label{fig:cosmo_paramsF}}\\
    \subfloat[Jerk parameter,  QCD]{\includegraphics[width=0.5\textwidth]{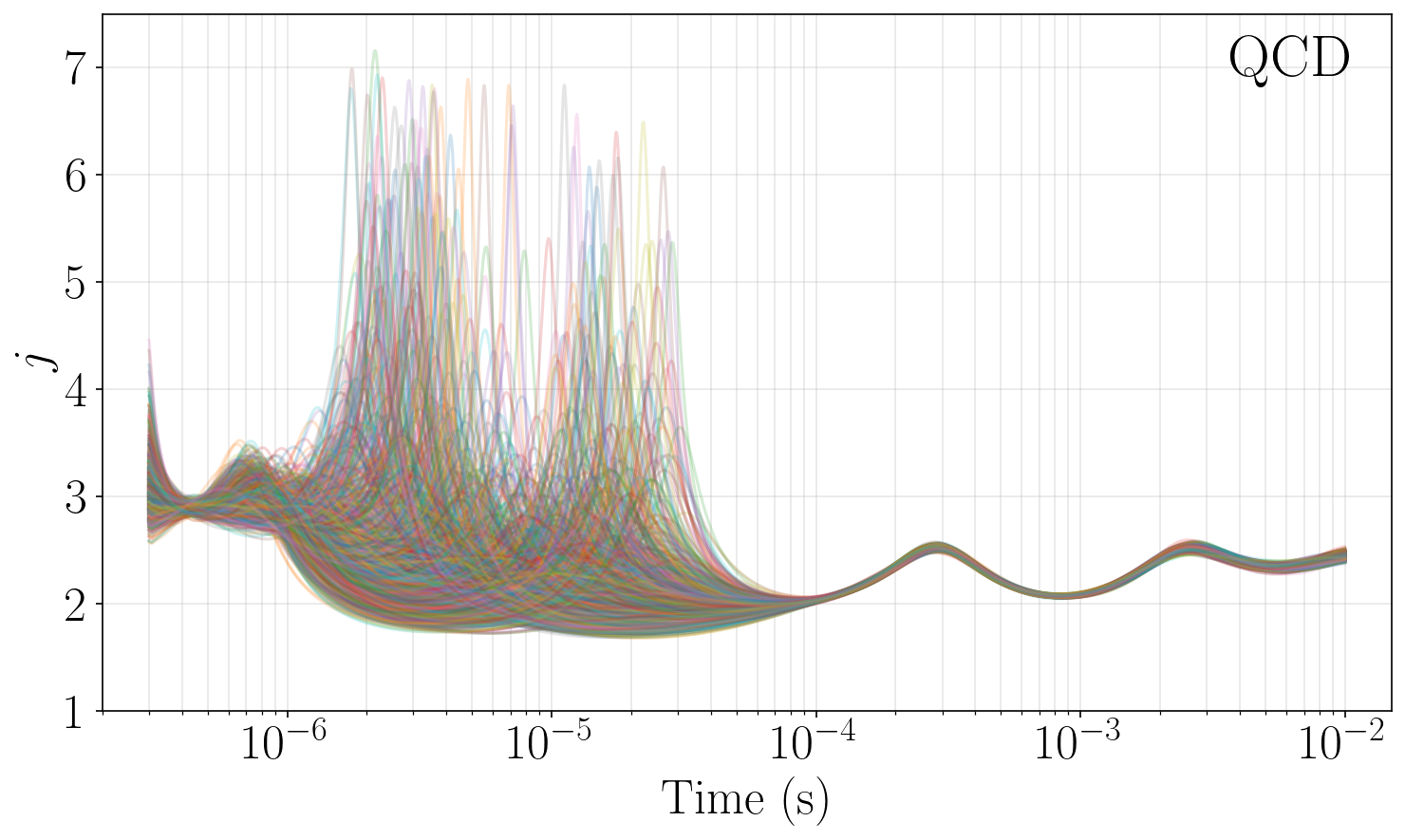}\label{fig:cosmo_paramsG}}
	\subfloat[Jerk parameter, QCD + EW]{\includegraphics[width=0.5\textwidth]{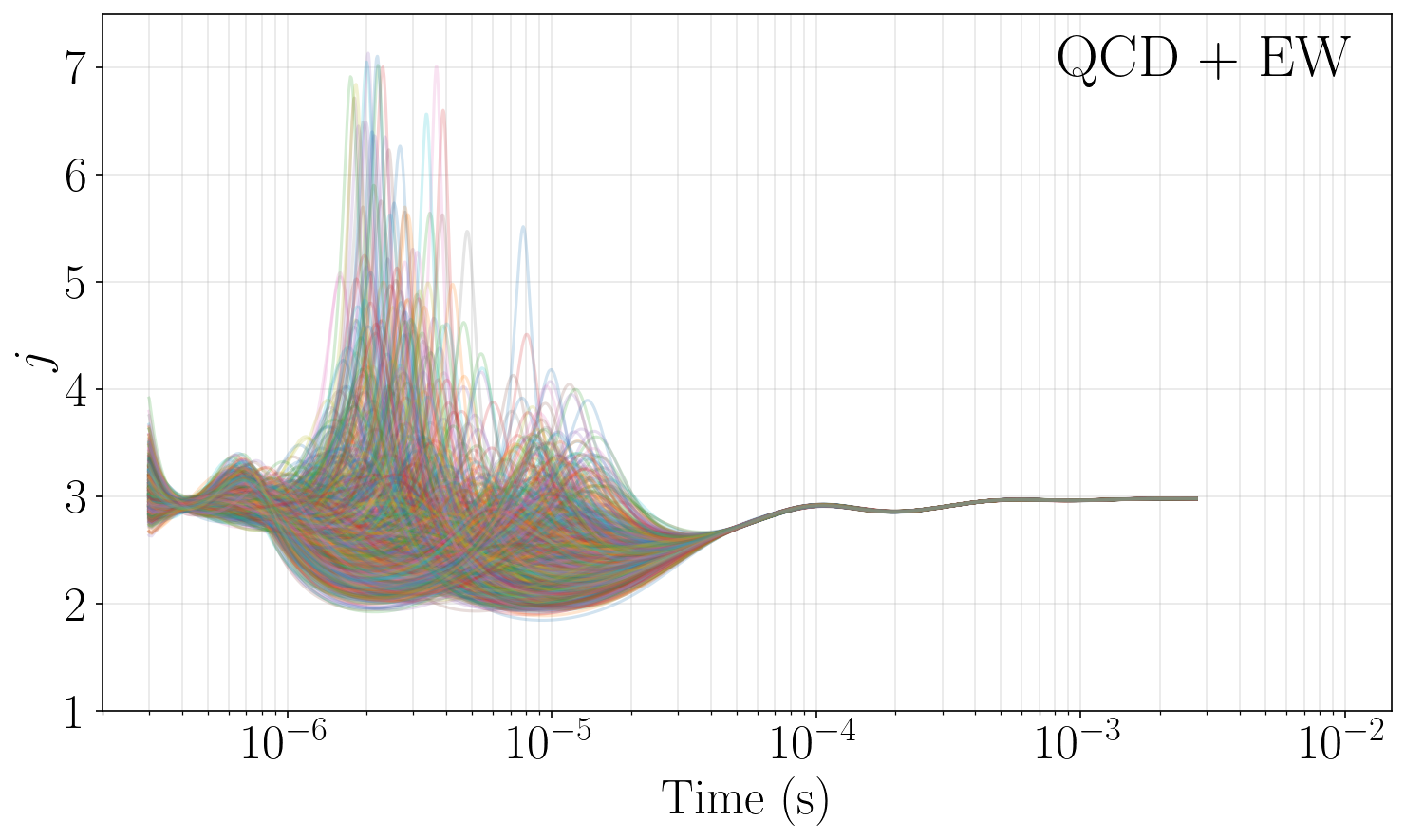}\label{fig:cosmo_paramsH}}
	\caption{Time evolution of cosmological parameters for 1000 GP-generated
           EoS.}
    \label{fig:cosmo_params}
\end{figure}

Initially, Fig.~\ref{fig:thermo_time} shows the time evolution of the temperature and $P/\varepsilon$. We initialize the evolution at $t_0 = 3\times 10^{-7}$~s with $T_0 = 944.7$~MeV and stop it when the temperature drops below 15~MeV. Our results indicate that the transition occurs for $t \ge 10^{-6}$ s, with the  temperature evolution being qualitatively similar in both cases. On the other hand, the approach of $P/\varepsilon$ to $1/3$ differs significantly. In particular, in the QCD-only scenario, $P/\varepsilon$ remains below $1/3$ for longer, delaying the onset of radiation-dominated expansion.
This delay directly impacts the expansion rate. As shown in
Fig.~\ref{fig:cosmo_params}, the Hubble parameter $H(t)$ is systematically
larger in the QCD-only case, reduced by approximately a factor of 2 at fixed
time when the EW sector is included. More strikingly, the higher-order
kinematic parameters exhibit qualitative differences: the deceleration
parameter $q(t)$ and jerk $j(t)$ display noticeable fluctuations around
$t \sim 10^{-5}$~s in the QCD-only case, which are strongly suppressed once
the EW contribution is included. This timescale corresponds precisely to
temperatures near $T \sim 0.2$~GeV, confirming that non-conformal QCD
dynamics induce transient features in the expansion history that are smoothed
out by the electroweak sector. This smoothing effect is consistent with the
expectation that additional relativistic degrees of freedom drive the universe
toward a more uniform radiation-dominated
expansion~\cite{Saikawa:2018,Borsanyi:2016ksw}.

\begin{figure}[t]
	\centering
	\subfloat[$\omega(T) = P/\varepsilon$]{\includegraphics[width=0.5\textwidth]{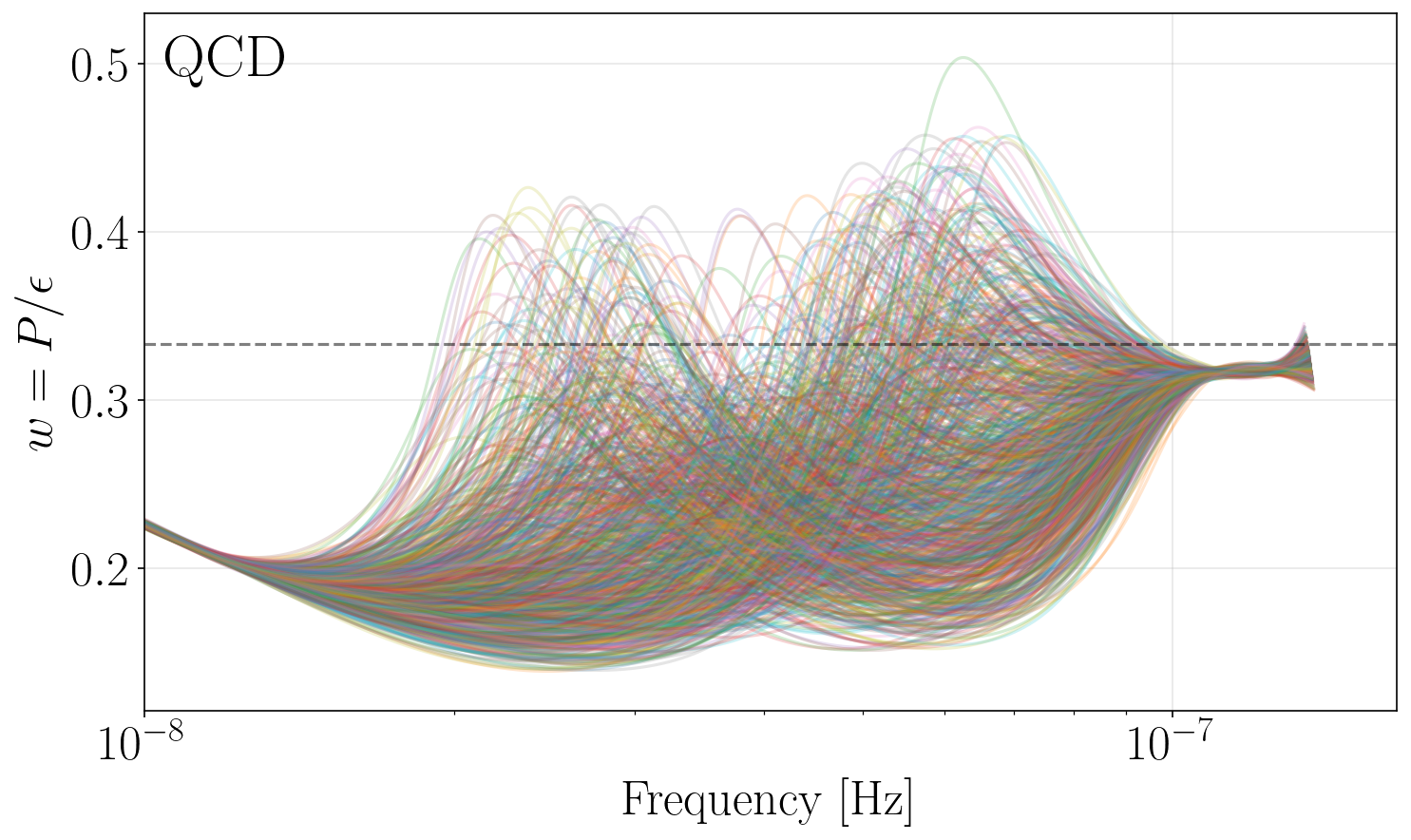}\label{fig:gwA}}
	\subfloat[$\omega(T) = P/\varepsilon$]{\includegraphics[width=0.5\textwidth]{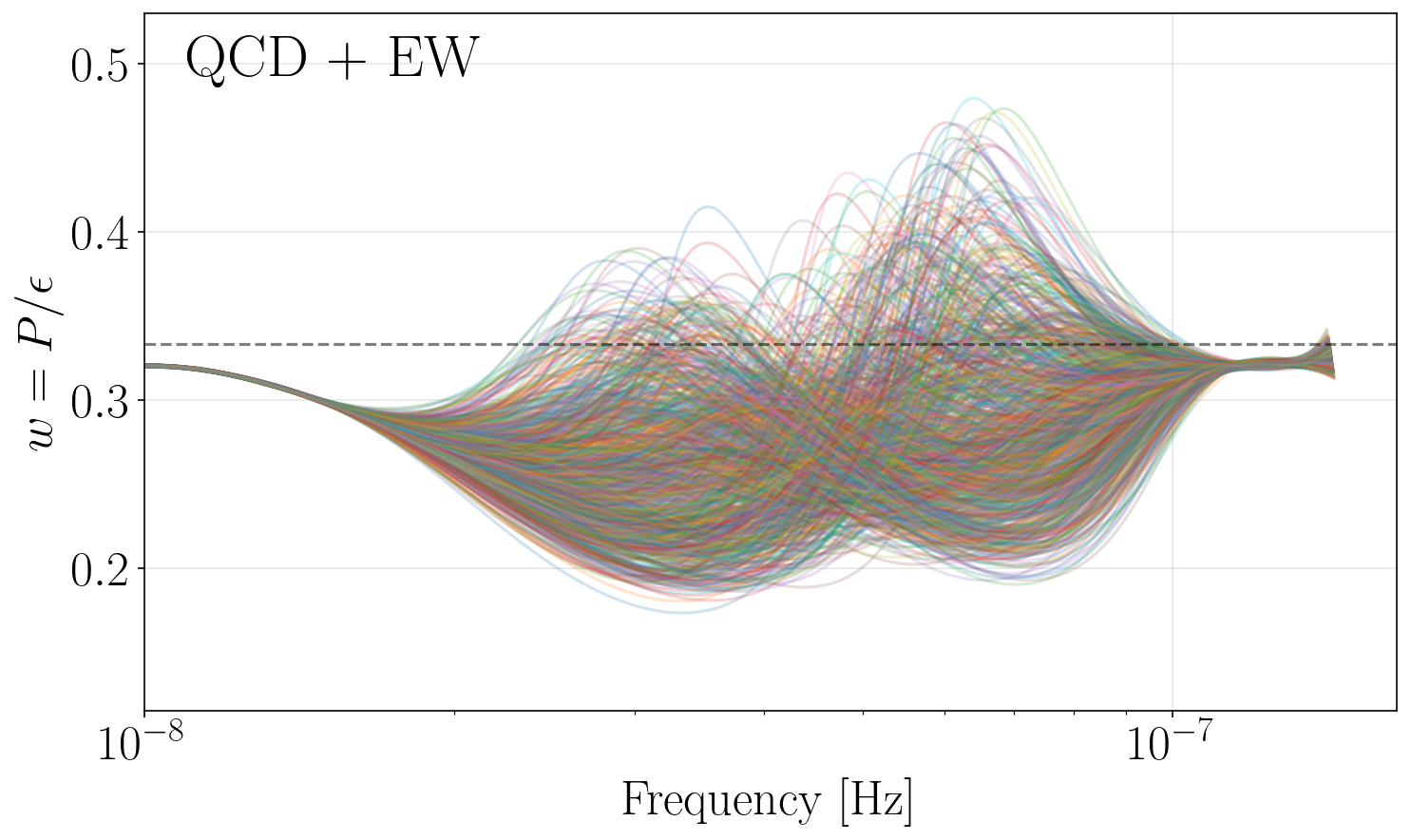}\label{fig:gwB}}\\
	\subfloat[Effective degrees of freedom $g_\mathrm{eff}(T)$]{\includegraphics[width=0.5\textwidth]{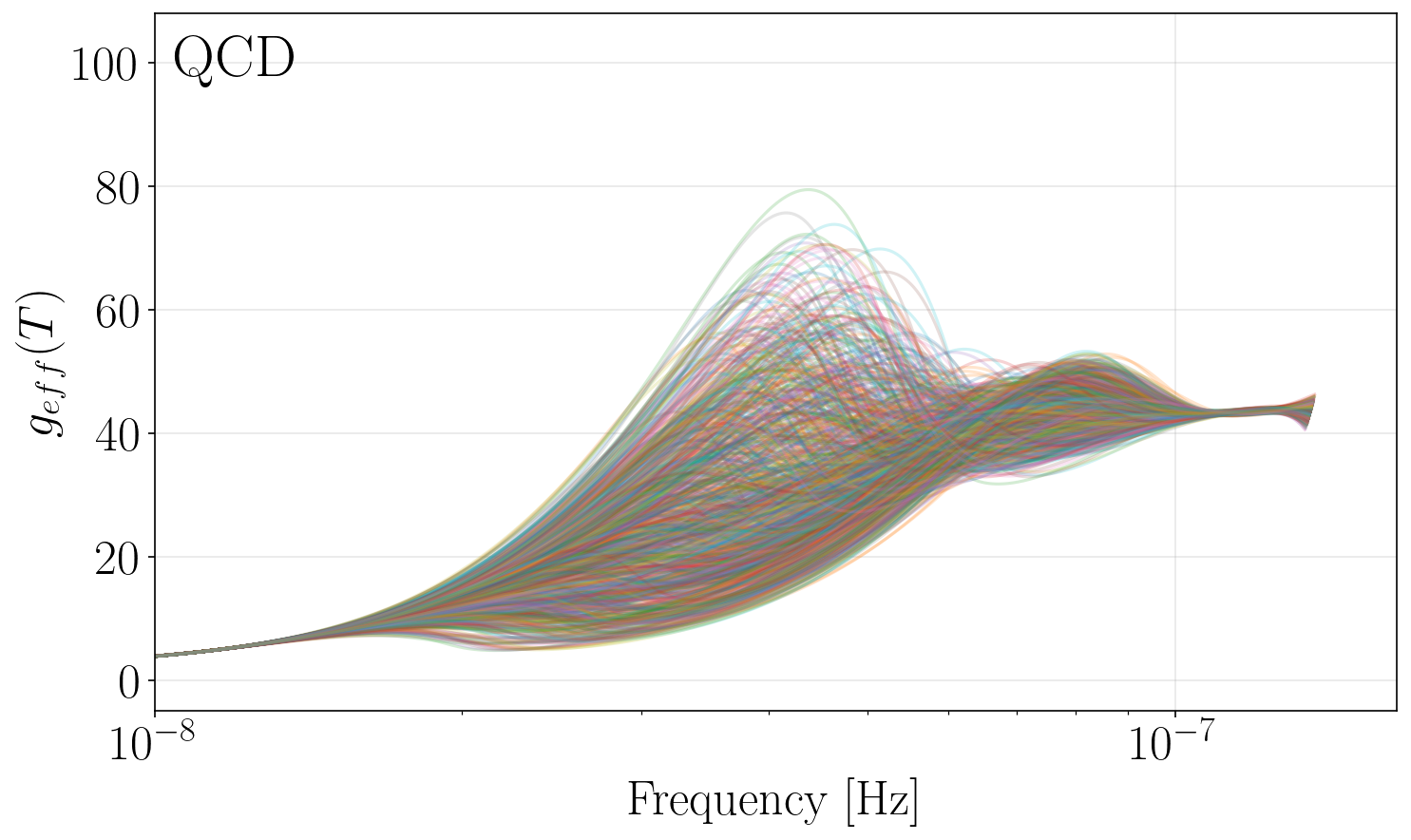}\label{fig:gwC}}
	\subfloat[Effective degrees of freedom $g_\mathrm{eff}(T)$]{\includegraphics[width=0.5\textwidth]{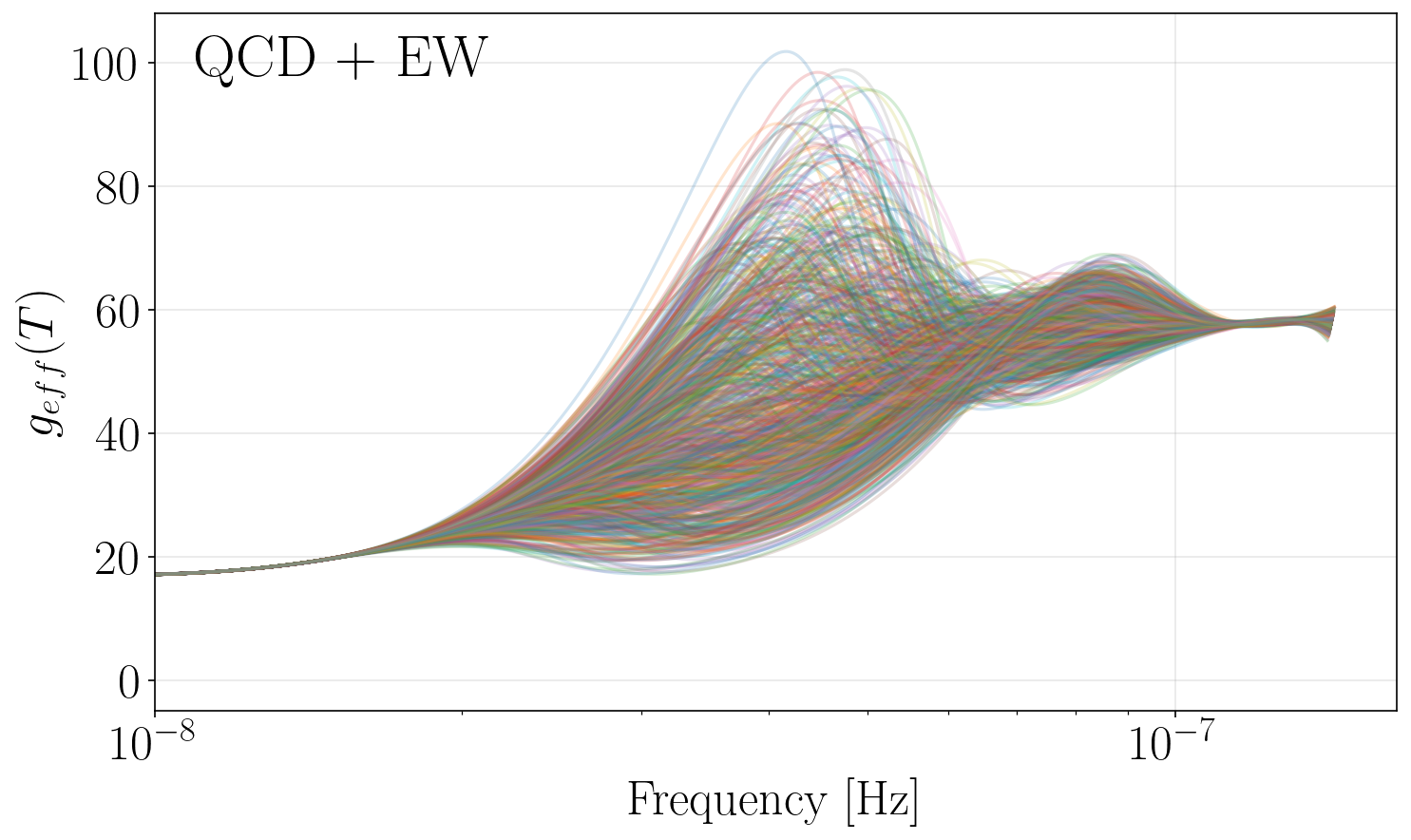}\label{fig:gwD}}\\
    \subfloat[Instantaneous spectral tilt $n_{gw}(f)$]{\includegraphics[width=0.5\textwidth]{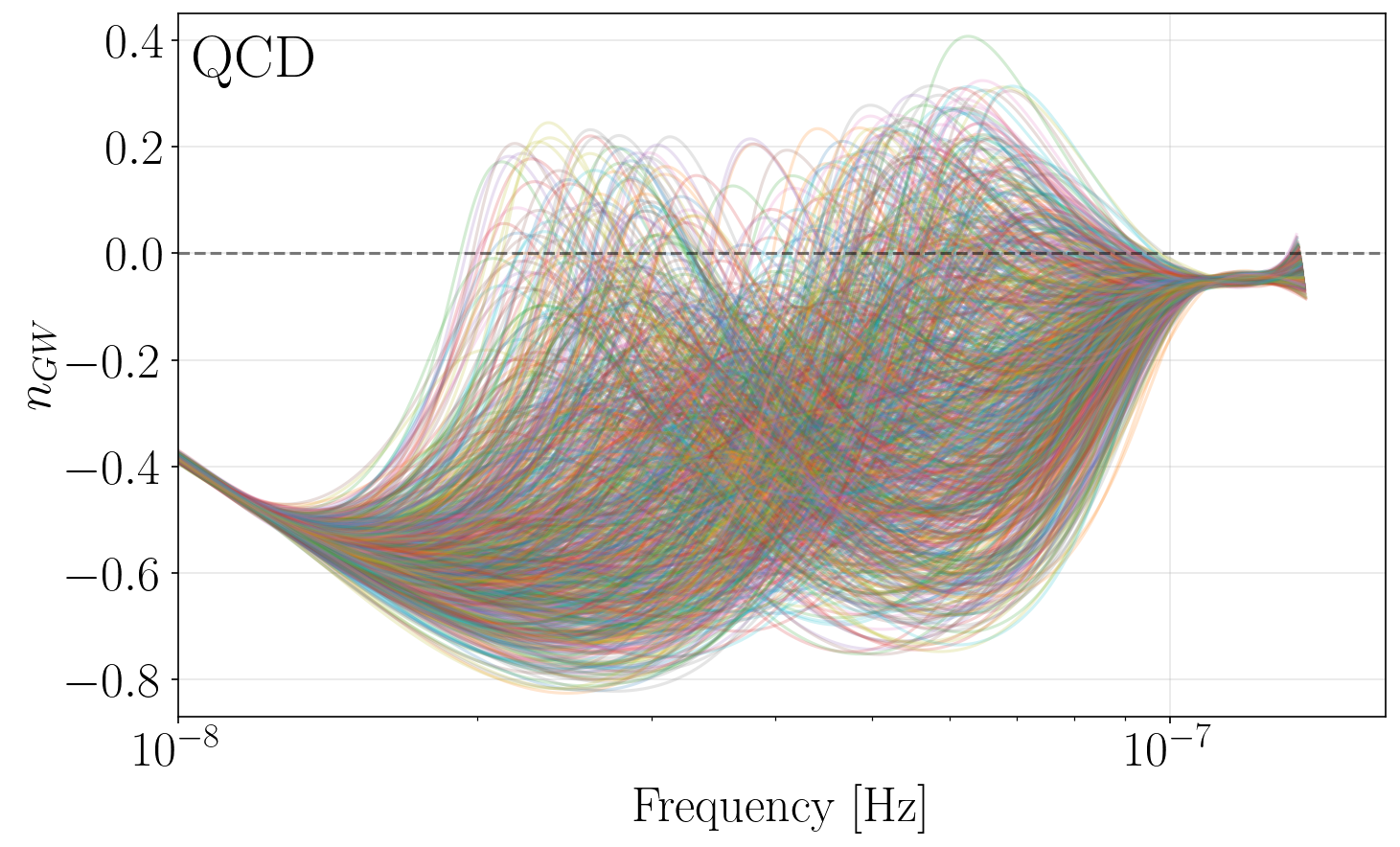}\label{fig:gwE}}
	\subfloat[Instantaneous spectral tilt $n_{gw}(f)$]{\includegraphics[width=0.5\textwidth]{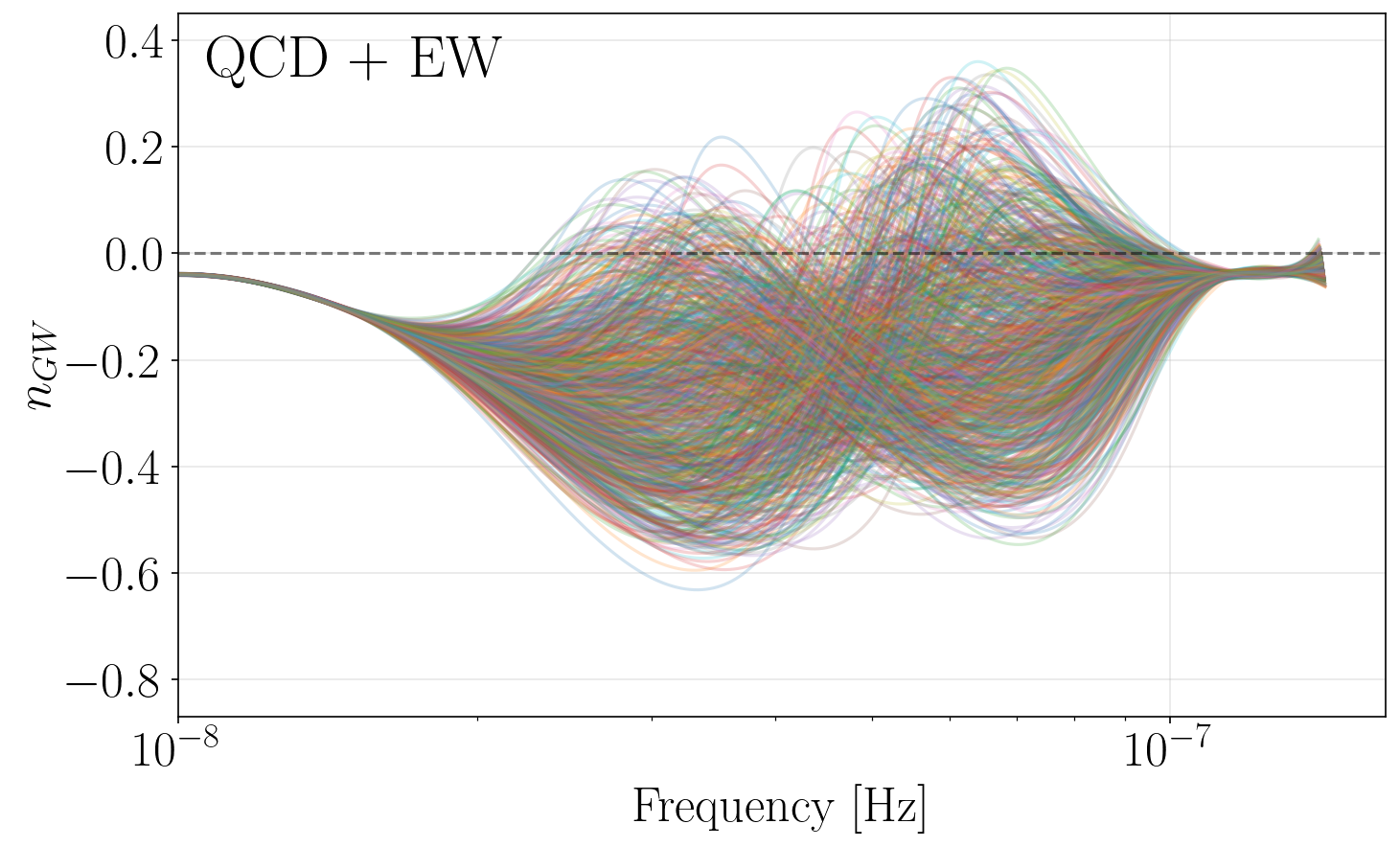}\label{fig:gwF}}\\
	\subfloat[Normalized GW energy density
           $\Omega_{gw}/\Omega_{gw}^{\rm rad}$]{\includegraphics[width=0.5\textwidth]{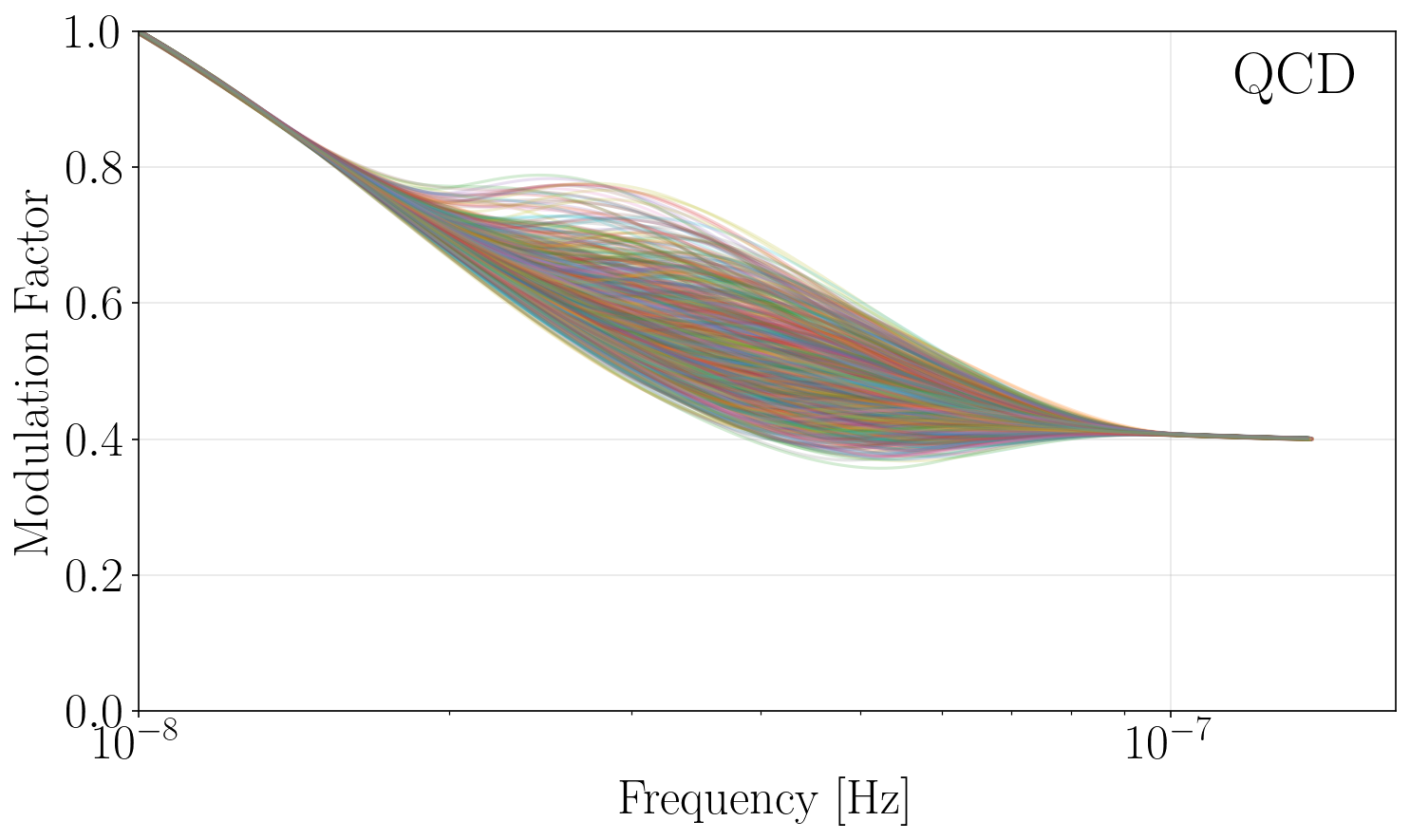}\label{fig:gwG}}
	\subfloat[Normalized GW energy density
           $\Omega_{gw}/\Omega_{gw}^{\rm rad}$]{\includegraphics[width=0.5\textwidth]{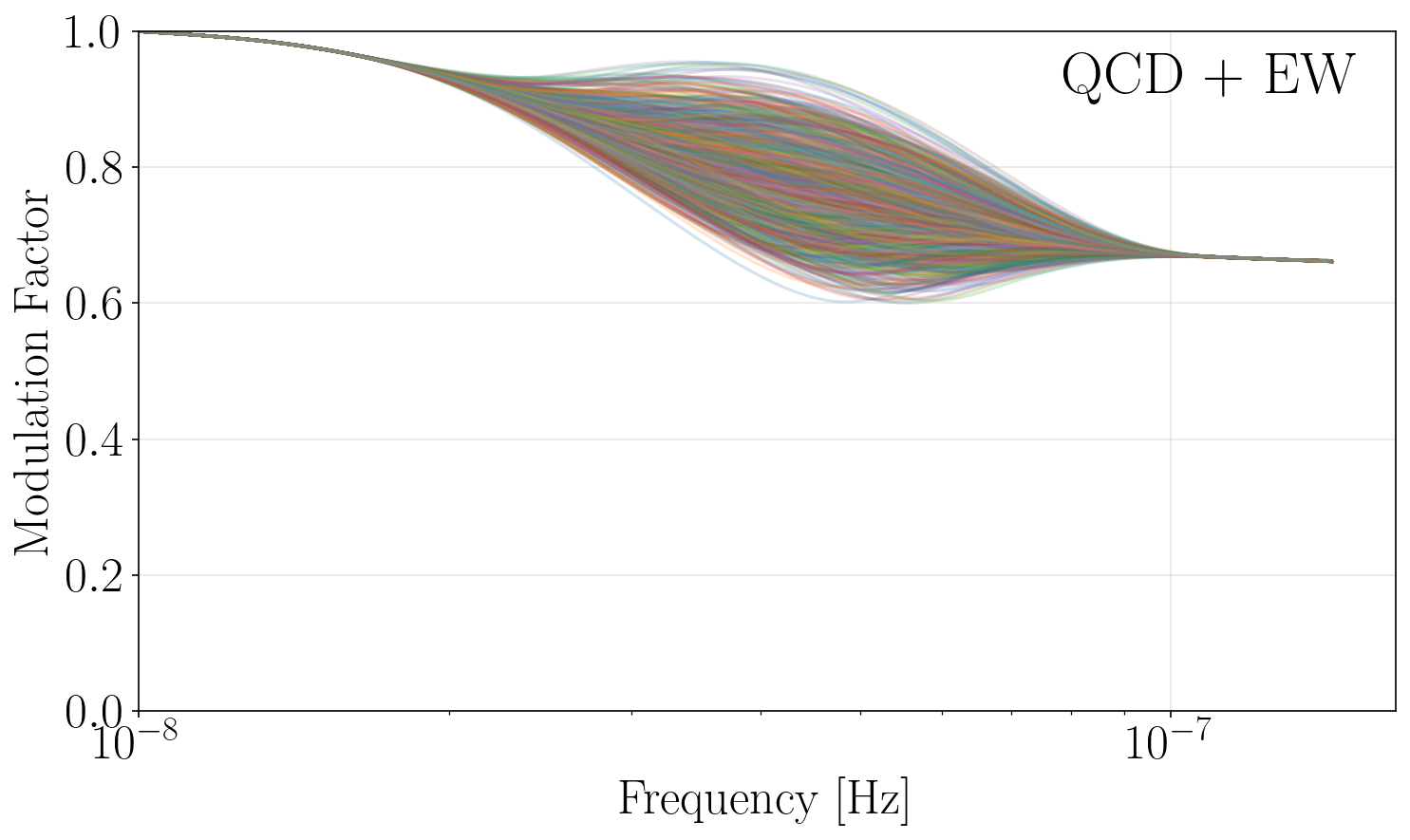}\label{fig:gwH}}
    \caption{Gravitational-wave observables as a function of frequency for 1000
           GP-generated EoS. }
    \label{fig:gw}
\end{figure}

Figure~\ref{fig:gw} shows the GW observables as a function of present-day frequency. The equation-of-state parameter $\omega(T)$  reaches the radiation limit $\omega = 1/3$ at significantly higher frequencies when the EW sector is included, reflecting the earlier onset of radiation domination. Moreover, the effective degrees of freedom $g_\mathrm{eff}(T)$  are correspondingly larger when the EW contribution is present, consistent with the additional particle content at high temperatures. The instantaneous spectral tilt $n_{gw}(f)$ approaches zero (scale invariance) at higher frequencies in the EW case. Without the EW contribution, a non-zero tilt persists over a broader frequency range, reflecting the extended period during which the universe remains non-conformal ($\omega < 1/3$). The most striking result appears in the normalized GW energy density $\Omega_{gw}/\Omega_{gw}^{\rm rad}$. In the QCD-only case, the spectrum exhibits a strong frequency dependence, rising significantly toward lower frequencies. When the EW sector is included, this behavior is replaced by a flat, suppressed plateau. This qualitative change demonstrates that electroweak degrees of freedom significantly suppress the imprint of QCD-induced features in the GW spectrum, driving it toward a universal radiation-dominated signature.

Taken together, these results establish a hierarchy in which electroweak
degrees of freedom dominate the high-temperature thermodynamic behavior and
significantly influence the effective equation of state. Their inclusion
reduces excursions toward the causality bound, smooths the expansion history,
and suppresses distinctive gravitational-wave signatures associated with the
QCD epoch. Consequently, extracting QCD-era physics from cosmological
observables---particularly from the stochastic gravitational-wave
background---requires careful modeling of the electroweak contribution to
avoid biased physical interpretations.
 
\section{Summary}
\label{sec:conclusions}
 
We have systematically investigated the impact of QCD equation of state and electroweak degrees of
freedom on the thermodynamic and cosmological evolution of the early universe,
employing Gaussian Process regression to generate ensembles of 1000 equations
of state built on the HotQCD lattice-QCD baseline. By comparing results with
and without the inclusion of the EW contribution, we have established a clear
hierarchy of effects that carries direct consequences for the interpretation
of cosmological observables.
 
At the level of thermodynamic quantities, the addition of the electroweak
sector significantly raises both the energy density and pressure, driving the
ratio $P/\varepsilon$ toward the radiation-dominated limit $1/3$ at higher
temperatures. Across the GP ensemble, this shift dominates over the intrinsic
spread arising from QCD modeling uncertainty, confirming that any EoS class
calibrated to lattice QCD at zero chemical potential will exhibit subdominant
inter-model differences once the EW contribution is consistently included.
 
The speed of sound provides the clearest diagnostic of this effect. Without
the EW contribution, a substantial fraction of GP samples approach the
causality bound $c_s^2 \to 1$ over a broad temperature interval around
$T \sim 0.2$~GeV, reflecting the non-conformal character of QCD matter in the
crossover region. The inclusion of the EW sector acts as a thermodynamic
regulator: by adding a nearly conformal component with
$P \approx \varepsilon/3$, it dilutes non-conformal QCD features and reduces
both the frequency and the temperature range over which large excursions in
$c_s^2$ occur.
 
These effects propagate directly into the cosmological expansion history.
The Hubble parameter is systematically reduced by approximately a factor of
two when the EW contribution is present, and the higher-order kinematic
parameters---the deceleration parameter $q(t)$ and the jerk $j(t)$---exhibit
qualitative differences. Transient fluctuations near $t \sim 10^{-5}$~s,
corresponding to temperatures around $T \sim 0.2$~GeV, are strongly
suppressed once the EW degrees of freedom are included, reflecting the
transition to a more uniform, radiation-dominated expansion driven by the
additional relativistic content.
 
The most observationally relevant consequences appear in the gravitational-wave
sector. The equation-of-state parameter $\omega(T)$ reaches the radiation
limit earlier in the EW case, causing the instantaneous spectral tilt
$n_{gw}(f)$ to approach zero at higher frequencies. Without the EW
contribution, the universe remains effectively softer over a broader frequency
range, imprinting a non-zero tilt that extends to lower frequencies. Most
strikingly, the normalized GW energy density
$\Omega_{gw}/\Omega_{gw}^{\rm rad}$ changes qualitatively: from a strongly
frequency-dependent, rising spectrum in the QCD-only case to a flat,
suppressed plateau when the EW sector is included.
 
Taken together, our results establish that the electroweak contribution is not
a minor correction but a dominant input to early-universe thermodynamics and
its cosmological consequences. Crucially, because all EoS considered here
share the same lattice QCD baseline, our conclusions are expected to hold for
any QCD-anchored EoS---including holographic models at zero chemical
potential---making them broadly applicable. For future efforts to extract
QCD-era physics from the stochastic gravitational-wave background, these
findings underscore the necessity of carefully modeling the electroweak sector:
omitting it introduces systematic biases in both the expansion history and the
predicted GW spectrum that could obscure or mimic genuine QCD signals.
 
\section*{Acknowledgments}

AVG acknowledges support from CNPq through the INCT-FNA grant 408419/2024-5, the Universal Grant 405458/2025-8 and the National Laboratory for Scientific Computing (LNCC/MCTI, Brazil), through the ambassador program (UFGD, subproject SADFT), for providing HPC resources of the SDumont supercomputer.
V.P.G. was partially supported by CNPq, FAPERGS and INCT-FNA (Process No. 408419/2024-5).

\bibliographystyle{apsrev4-2}


\end{document}